\documentclass[12pt,a4paper]{article}
\usepackage[margin=2cm]{geometry}
\usepackage{psfrag}
\usepackage{amsmath}
\usepackage{amsfonts}
\usepackage{amssymb}
\usepackage{graphicx}
\usepackage{pstricks}
\usepackage{hyperref}

\usepackage{lscape,epsf}

\newcommand{\MET}{\mbox{$\not\hspace{-0.11cm}E_T$}}
\newcommand\Trule{\rule{0pt}{2.6ex}}
\newcommand\Brule{\rule[-1.2ex]{0pt}{0pt}}

\def\Fc{{\cal{F}}}

\def\PFGstripminus-#1{#1}%
\def\PFGshift(#1,#2)#3{\raisebox{#2}[\height][\depth]{\hbox{%
  \ifdim#1<0pt\kern#1 #3\kern\PFGstripminus#1\else\kern#1 #3\kern-#1\fi}}}%

\newcommand{\ME}[3]{\langle #1 | #2 | #3 \rangle}

\newcommand{\GeV}{{\rm\ GeV}}
\newcommand{\MeV}{{\rm\ MeV}}

\newcommand{\lag}{{\cal L}}

\renewcommand{\theequation}{\thesection.\arabic{equation}}

\newcommand{\TeV}{{\rm\ TeV}}
\newcommand{\tr}{{\rm tr \ }}
\def\ds{\displaystyle}
\def\bea{\begin{array}{c}}
\def\ea{\end{array}}
\def\be{\begin{equation}\bea\ds}
\def\ee{\ea\end{equation}}

\begin{document}

\begin{titlepage}

\begin{flushright}
{RUNHETC-2009-25}\\
\end{flushright}

\begin{center}
{\large {\bf Pure-glue hidden valleys through the Higgs portal
}} \\
\vskip0.5cm {Jos\'e~Juknevich}
\vskip1cm
{\itshape Department of Physics and Astronomy,\\
Rutgers, 136 Frelinghuysen Rd, Piscataway, NJ 08854, USA\\

}
\end{center}

\vspace{0.5cm}
\begin{abstract}
We consider the possibility that the Higgs boson can act as a link to a hidden sector in the context of pure-glue hidden valley models.
In these models the standard model is weakly coupled, through loops of heavy messengers fields, to a hidden sector whose low energy dynamics is described by a pure-Yang-Mills theory. Such a hidden sector contains several metastable hidden glueballs.  In this work we shall extend earlier results on hidden valleys to include couplings of the messengers to the standard model Higgs sector. The  effective interactions at one-loop couple the hidden gluons to the standard model particles through the Higgs sector. These couplings in turn induce hidden glueball decays to fermion pairs, or cascade decays with multiple Higgs emission.
The  presence of effective operators of different mass dimensions, often competing with each other, together with a great diversity
of states, leads to a great variability in the lifetimes and decay modes of the hidden glueballs.
We find that most of the operators considered in this paper are not heavily constrained by precision electroweak physics, therefore leaving
plenty of room in the parameter space to be explored by the future experiments at the LHC. 
\end{abstract}

\end{titlepage}

\tableofcontents


\section{Introduction}
\setcounter{equation}{0}
In view of the construction of the Large Hadron Collider (LHC) and its detectors, there have been major efforts to develop ideas for what physics may lie beyond the standard model (SM). This work is often motivated by a number of problems that still remain unsolved such as the hierarchy problem, baryogenesis and the origin of dark matter, among others.  Nevertheless,   it is also possible that  new physics is unrelated to  the above phenomena and may show up in a way that is not predicted by existing theories. To ensure that such a physics does not go undiscovered requires precise understanding of how new physics will reveal itself in the next particle-physics experiments.  Therefore, it is prudent to explore other approaches to studying new theories, with an emphasis in models with varied experimental signatures.

The hidden valley scenario \cite{SZ,HV2,HV3,HVun,HVWis,hvstudy1} has been proposed as an interesting example of physics beyond the SM that can profoundly affect future experiments at the LHC. A hidden valley refers to any new sector (``v-sector") that is added to the SM with its own particles and interactions, neutral under the SM. A mass gap ensures that not all the particles in the v-sector decay to extremely-light, invisible particles. However, they can decay through the very weak interactions to the SM, often producing visible signals. Hidden valleys can lead to interesting and unexpected collider phenomenology, including  hidden states decaying  macroscopic distances away from the primary interaction vertex, or unusual   final states  with a high multiplicity of SM particles; see for example  \cite{SZ,HV2,hvstudy1}. 

Hidden valleys have arisen in bottom-up models such as the twin-Higgs
and folded-SUSY models \cite{TwinHiggs,FoldedSUSY} that
attempt to address the hierarchy problem, and in a recent attempt to
explain the various anomalies in dark-matter searches
\cite{hvdarkmatter}.  They are also motivated by string compactifications: hidden sectors that are candidate hidden valleys often arise
in many phenomenological string theory models which aim to come as close as possible to the minimal supersymmetric standard model, but  typically have extra vector-like matter and extra gauge groups, see for example \cite{string hv}.

So far various realizations of hidden valleys have been proposed, which can be generally categorized according to the matter content in the v-sector and the mechanism which communicates the interactions between the two sectors. 
An interesting   class of hidden valleys is obtained by adding to the SM a v-sector which at low energy is a
pure-Yang-Mills theory with gauge group $G_v$, a theory that has its own gluons
(``v-gluons'') and their  bound states (``v-glueballs'') ~\cite{Juknevich:2009ji}. In addition, the model contains some heavy matter fields $X_r$, charged under both $G_v$ and SM gauge group, mediating the interaction between the SM and the v-sector. Such states are called quirks in analogy to traditional quarks\footnote{Quirks were considered long ago \cite{Okun,quinn}; some of their very interesting dynamics have been studied in \cite{SZ,KLN}.}. Without additional massless flavors in the v-sector, the gauge group $G_v$ is confining and  the dynamics of  the v-sector  will develop a mass gap, slightly larger than the confining scale $\Lambda_v$ where the $G_v$ coupling gets strong. 
This scenario  easily arises in many supersymmetric models with extended gauge groups beyond the standard model; for example, in many supersymmetric v-sectors, supersymmetry breaking  may lead to large masses for all matter fields.
  
\begin{figure}[ht]
\begin{center}
\epsfxsize=12cm \epsffile{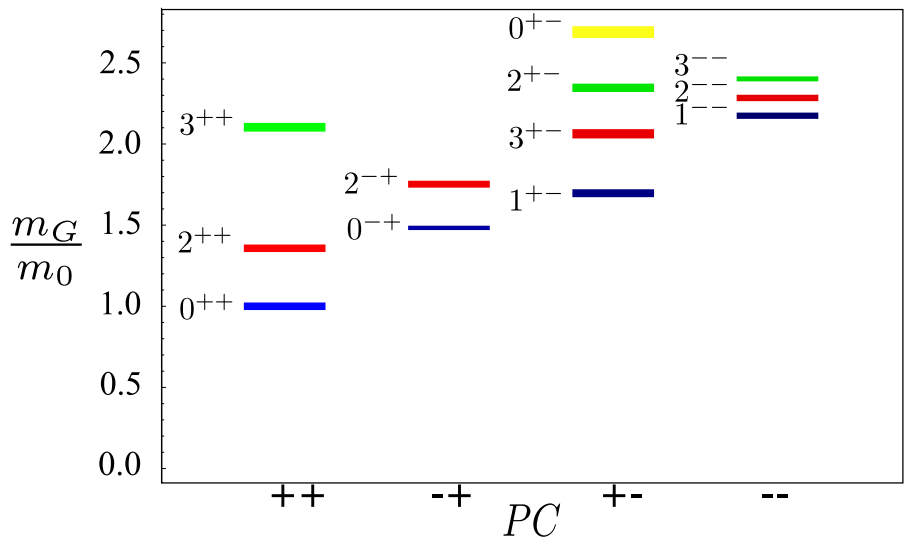}
\end{center}
\vspace{-0.6cm}
\caption{ Spectrum of stable glueballs in pure glue $SU(3)$ theory \cite{Morningstar}. Masses are shown in units of the lightest v-glueball mass $m_0$.}
\label{spectrum}
\end{figure}

The spectrum of a pure Yang-Mills theory has been studied on the lattice for the case of a  $SU(3)$ gauge group \cite{Morningstar}. These results are shown in figure \ref{spectrum}. The spectrum of stable bound states includes several glueballs with different quantum numbers $J^{PC}$. Some of these glueballs
 have been studied in some detail in various contexts; see for
example~\cite{Morningstar2,Shifman scalar,Shifman
pseudoscalar,MITbagmodel,glueballs,bagmodel,Loan} and a recent review~\cite{glueballreview}. The masses of the v-glueballs are determined by the confinement scale $\Lambda_v$ or, equivalently, the string tension $\sigma$ through the relationship $m_{0^{++}}\approx 7 \Lambda_v\approx 3.7\sigma$. For general  $SU(n_v)$ gauge group, the leading large $n_v$ corrections to the glueball masses are of order $1/n_v^2$.  Therefore,  we expect that the mass ratios in the general $SU(n_v)$ case will not differ substantially from the mass ratios shown in figure \ref{spectrum} for $n_v=3$. All of the v-glueballs shown are stable against decays to the other v-glueballs, due to kinematics and/or conserved quantum numbers.

The v-glueballs are non-interacting with the SM particles, except for higher dimension operators in the effective Lagrangian induced by the heavy mediators $X_r$.
 In this work, we will focus on  the possibility of operators between the v-sector and the Higgs sector (also referred to as the Higgs portal),
\be\label{portal}
\frac{1}{M^{D-4}}\, {\cal O}_s^{D-d}(H^\dagger, H) \, {\cal O}_v^{(d)}
\ee
where $H$ is the standard model Higgs doublet; $M$ is a heavy mass scale, associated with the masses of the heavy mediator fields.   Here we
have split the dimension-$D$ operator into a
Standard-Model part ${\cal O}_s^{(D-d)}$ of dimension $D-d$ and a
hidden-valley part ${\cal O}_v^{(d)}$ of dimension $d$. 
The  ${\cal O}_v^{(d)}$ are constructed from  gauge invariant combinations of v-gluon fields such as $ \tr  {\cal F}^2 $ or $ \tr  {\cal F}^3 $, while the ${\cal O}_s^{(D-d)}$ involve gauge invariant operators built out of the Higgs field, such as $H^\dagger H$ or  $H^\dagger D_\mu H$. We will see that the lowest order term is given by the dimension-six operators of the form $H^\dagger H \,\tr {\cal F}^2$. With some exceptions to be discussed below, most of  the v-glueballs in figure \ref{spectrum} can decay via the  Higgs portal interaction (\ref{portal}), with a strong dependence of the lifetimes on the confinement scale $\Lambda_v$ and the mass scale $M$.

Loops of the  heavy particles also induce dimension-eight effective interactions coupling the v-gluons to the standard model gauge bosons, either of the form $\tr{F}_{i}^2 \,\tr {\cal F}^2 $ or ${F_1} \,\tr {\cal F}^3$ where $F_i$ (${\cal F}$) is the field strength tensor for standard model (hidden-valley) gauge bosons. The field strength tensors are contracted according to different irreducible representations of the Lorentz group.  A detailed study of the phenomenology of these operators in the context of hidden valleys was carried out in  ~\cite{Juknevich:2009ji}. In this case the v-glueball widths are dominated by decays into SM gauge-boson pairs, or radiative decays to another v-glueball and a photon (or to a lesser extent a Z boson), leading to jet and photon rich final states.

In this paper, we will first  extend the results of  ~\cite{Juknevich:2009ji} on v-glueball decays in pure-Yang-Mills theory to allow for higher dimension interactions of the form (\ref{portal}). We will need to compute the effective Lagrangian coupling the v-gluons to the Higgs sector. Then we will use it to find formulas for the decay widths  of the states in the spectrum shown in figure \ref{spectrum}. Specifically, we will find that the $0^{++}$ state can decay via the Higgs portal to standard model particles with branching ratios which are determined by the couplings of the standard model Higgs boson. Other states can decay to a lighter state by emission of a Higgs boson. For v-glueball states which can decay either through dimension-eight operators or through the emission of a Higgs boson the situation is rather involved, with the relative branching ratios depending on the various parameters and the  unknown v-glueball matrix elements. 

Our primary motivation here is in the case where the lifetimes are short enough that at least a few v-glueball decays can be observed at the LHC detectors. This typically requires the lifetimes of the hidden particles to be shorter than a few micro-seconds, if the production cross-section is substantial.
The key point in our scenario is that with different operators describing the decays, often competing with each other, we typically find a large spread in the lifetimes of the v-glueballs, even after all the parameters have been fixed. Then there is no necessity to adjust any parameter to obtain short lifetimes.

 Incidentally, our results are also relevant for studies of dark matter, either in the case of self-interacting dark matter ~\cite{Faraggi:2000pv}, where the v-glueballs would be the dark matter candidates,  or indirectly, in a recent attempt to study scenarios of dark matter with novel signatures~\cite{Falkowski:2009yz,Kribs:2009fy}.

This work is organized as follows. In section 2, we introduce our model and describe the effective interactions coupling the two sectors. We also classify the matrix elements. Section 3 presents our computation of the decay modes. Section 4 is devoted to a summary of the different  experimental constraints  on the parameters of the model. Then in section 5 we present our numerical estimates for the branching ratios. Possible generalizations of the model are described in section 6. Finally, we conclude in section 7 with a brief summary of our results and some comments. Additional computations appear in the appendix.

\section{The model and the effective action}
\setcounter{equation}{0}
We first set up our framework and conventions.
The generic scenario of hidden valleys with a  pure-gauge hidden sector can be characterized as follows \cite{SZ,Juknevich:2009ji},
\begin{itemize}
\item The SM is extended by the addition of an extra $SU(n_v)$ gauge group, with a mass scale scale $m_0$ in the $1-1000 \GeV$ range. We will refer to this sector as the hidden valley, or briefly the v-sector. There are no light flavours in the v-sector, so after confinement the lightest states in the  spectrum are bound states of v-gluons, or v-glueballs.
\item There are heavy vector-like particles $X_r$ (``mediators'') that couple the v-sector very weakly to the visible sector  at low energies. For definiteness, we will take the  $X_r$ to transform as a fundamental representation of $SU(n_v)$ and in complete $SU(5)$ representations of the standard model. We label the fields and their masses as shown in table \ref{tab reps}. Also, for convenience, we define dimensionless parameters
 $
\rho_r=m_r/M
$ where $M$ is an arbitrarily chosen mass scale, usually taken as the mass of the lightest $X_r$ particle\footnote{In this work, we normalize  hypercharge as $Y= Q-T_3$, where $T_3$ is the third component of  weak isospin.}.
\end{itemize}
\begin{table}[h]
\begin{center}
\begin{tabular}[c]{|c|c|c|c|c|c|}\hline
\Trule\Brule Field & $SU(3)$ &$SU(2)$& $U(1)$
& $SU(n_v)$ & Mass
\\ \hline
\Trule\Brule $X_{\bar d}$ & ${\bf 3}$ & ${\bf  1}$  & $\frac13$
&  ${\bf n_v}$  & $m_{\bar d}$
\\ \hline
\Trule\Brule $  X_\ell $ & ${\bf 1}$ & ${\bf 2}$  & ${-\frac12}$
&  ${\bf n_v}$ & $m_{\ell}$
\\ \hline \Trule\Brule
$X_{\bar u} $ & $\bf \bar 3$
& $\bf 1$  & $-\frac23$
& $\bf n_v $ &$m_{\bar u}$
 \\  \hline \Trule\Brule
$X_q$ & $\bf 3$ & $\bf 2$  & $\frac16$
& $\bf n_v$ &$m_q$
\\ \hline \Trule\Brule
$X_e$ & ${\bf 1}$ & ${\bf 1}$  & ${\bf  1}$
&  ${\bf n_v}$ & $m_e$
\\ \hline
\end{tabular}
\end{center}
\caption{\small The new fermions $X_r$ that couple the hidden valley
sector to the SM sector.}
\label{tab reps}
\end{table}

In addition, we assume that the mediators $X_r$ can get part of their mass from electroweak symmetry breaking through Yukawa-like interactions. For concreteness, we consider the following Lagrangian (in four-component Dirac notation),
 \be
\lag_{mass}=\sum_{r=\bar d,l,\bar u,q,\bar e} m_r \bar X_r X_r +\left(y_l \bar X_l H X_{\bar e} + y_u  \bar X_q \tilde H X_{ u}+ y_d\bar X_q H X_{ d}+h.c.\right) \label{yuk}
\ee
where $\tilde H = \epsilon \cdot H^\dagger$.
For the moment, we will restrict ourselves to the case in which the Yukawa couplings conserve both $C$ and $P$ independently.
In general one or more of the  Yukawa couplings could be $CP$-violating. This $CP$ violation can contribute new terms to the effective action and can induce additional v-glueball decays, as  will be discussed in section 6.

There are several constraints on the couplings $y_r$ and the mass scale $M$ from precision electroweak measurements. These constraints will be discussed in some detail in section 4. We require that $y_r \lesssim 1$ and $M\gtrsim 250\GeV$ in order to avoid potentially dangerous corrections to   precision electroweak bounds. A high mass scale is in any case  necessary to avoid current experimental bounds from collider searches for extra  particles.

\begin{figure}[ht]
\begin{center}
\epsfxsize=12cm \epsffile{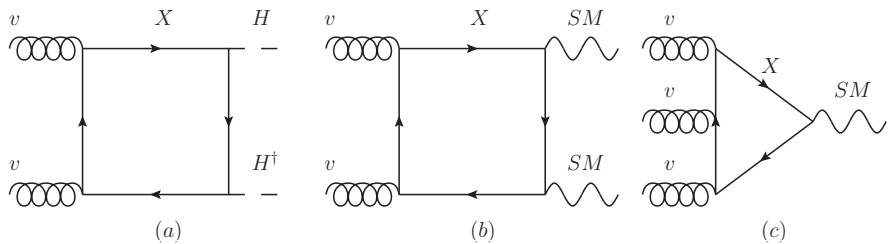}
\end{center}
\caption{Diagrams contributing to the effective action}
\label{loops}
\end{figure}

We are concerned with the low-energy effective theory of the model described above. 
The  effective interaction that couples the v-gluons and v-glueballs to the SM particles is induced through a loop of $X$ particles.  The coupling between the Higgs sector and the v-sector to non-vanishing leading order in $1/M$ at low energies arises from a diagram with four external lines, as depicted in  figure \ref{loops}a. This gives rise to the dimension-six operator
\be\label{efflag}
{\cal L}^{(6)}= \frac{\alpha_v \,y^2}{ 3 \pi\, M^2}\,   H^\dagger H \,\tr {\cal F}_{\mu\nu} {\cal F}^{\mu\nu}
\ee
where $\alpha_v$ is the $SU(n_v)$ coupling, ${\cal F}_{\mu\nu}$ is the v-gluon field strength tensor\footnote{Here we represent the v-gluon fields as ${\cal F}_{\mu\nu}={\cal F}_{\mu\nu}^a T^a$, where $T^a$ denote the generators of the $SU(n_v)$ algebra with a common normalization $\tr T^aT^b=\frac12 \delta^{ab}$. }.
The coefficient $y$ depends on the mass ratios of the heavy particles from table \ref{tab reps} and their couplings to the Higgs field. It is given by
\be\label{y}
y^2 \equiv \frac{  y_l^2 }{\rho_l\rho_{e}}+\frac{3 y_d^2}{ \rho_q \rho_{\bar d}}+\frac{3  y_u^2}{\rho_q \rho_{\bar u}}.
\ee
The computation of ($\ref{efflag}$) is presented in appendix A. Note that (\ref{efflag})-(\ref{y}) are strictly valid in the $ \beta_r \equiv y_r^2 v_H^2/2M^2 \ll 1$ limit; otherwise, the corrections to these equations can be readily obtained by the substitutions $\rho_r \rho_{r'} \to \rho_r \rho_{r'} -\beta_r$ in (\ref{y}).
Once the Higgs gets an expectation value, the following interaction terms are induced between the v-sector and the physical standard model Higgs boson ($h$),
\be\label{efflag-exp}
{\cal L}^{(6)}= \frac{\alpha_v \,y^2}{3\pi M^2} \, v_H\,h \,\tr {\cal F}_{\mu\nu} {\cal F}^{\mu\nu}+ \frac{\alpha_v \,y^2}{3\pi M^2}\,   \frac{h^2}{2} \,\tr {\cal F}_{\mu\nu} {\cal F}^{\mu\nu}
\ee
where $v_H=246\GeV$ is the Higgs vacuum expectation value\footnote{In (\ref{efflag-exp}), we have not included a term which is proportional to the v-gluon kinetic term. This term can be removed by  redefining the v-gluon coupling.}.

While dimension-six operators are only suppressed by two powers of the mass scale $M$, further suppression can arise if the operators are suppressed by small coupling $y$. This requires us to consider  effective operators of dimension higher than six as well.

The next operators have $D=8$, and they describe the coupling of the two v-gluons to two SM gauge bosons (figure \ref{loops}b), either gluons ($g$), weak bosons ($W$ and $Z$) or photons ($\gamma$), as well as the fusion of three v-gluons into a $\gamma$ or $Z$ (figure \ref{loops}c).  Within
the SM, effective two gluon - two photon, four gluon, and
three gluon - photon vertices can be found in \cite{Shifman
charmonium}, \cite{Strassler:1992nc} and \cite{Groote}
respectively.  They were also extensively discussed in a preceding work ~\cite{Juknevich:2009ji}. The pure-gauge effective Lagrangian linking
the SM sector with the v-sector reads  (for our conventions see Appendix B)
\begin{multline}
\label{eff_lag full2}
 {\cal L}^{{\rm (8)}}= \frac{g_v^2}{(4\pi)^2M^4}\left[g_1^2\chi_1 {B}^{\mu\nu}{B}^{\rho\sigma} +
g_2^2\chi_2 \tr{F}^{\mu\nu}{F}^{\rho\sigma} +
g_3^2\chi_3 \tr{G}^{\mu\nu}{G}^{\rho\sigma} \right]\\
 \ \ \ \ \ \ \ \ \ \ \ \ \ \ \ \ \ \ \ \ \times
\left(
  \frac{1}{60} \,S\, g_{\mu\rho}g_{\nu\sigma}
  + \frac{1}{45} \,P\, \epsilon_{\mu\nu\rho\sigma}
 + \frac{11}{45} \,T_{\mu\rho}g_{\nu\sigma}\, - \frac{1}{30}  \,L_{\mu\nu\rho\sigma}\,  \right) \\ +
 \frac{g_v^3g_1}{(4\pi)^2M^4}\,\chi\left(\frac{14}{45}\,
 B^{\mu\nu}\Omega^{(1)}_{\mu\nu} - \frac{1}{9}\,
 B^{\mu\nu}\Omega^{(2)}_{\mu\nu}\right).
\end{multline}
Here we  define $F_{\mu\nu}^1\equiv
B_{\mu\nu}$, $F_{\mu\nu}^2\equiv F_{\mu\nu}$ and $F_{\mu\nu}^3\equiv
G_{\mu\nu}$, which are the field tensors of the $U(1)_Y$, $SU(2)$ and
$SU(3)$ SM gauge groups. We denote their couplings $g_{i}$, $i=1,2,3$,
while $g_v$ is the coupling of the new group $SU(n_v)$.
The coefficients $\chi_i$ and $\chi$ encode the masses of the heavy particles from table \ref{ratios} and their couplings  to the SM gauge bosons and the Higgs doublet. They are summarized in table \ref{xi}.
  The $d=4$ operators ($S$, $P$, $T$, $L$) involve  bilinear $\tr {\cal F F}$ operators whereas the $d=6$ operators ($\Omega^{(1)}$, $\Omega^{(2)}$) involves trilinears of the form  $\tr {\cal F F F}$ (see Appendix B).

 \begin{table}[ht]\label{ratios}
\begin{center}
 \begin{tabular}[c]{|c|c|}\hline \Trule\Brule
\  &$\chi$ , $\chi_i$
\\ \hline \Trule\Brule
$\chi_1$ & $ \frac{1}{ 3 \rho_{\bar d}^4}+\frac{1}{2 \rho_{ l}^4} +\frac{4}{3 \rho_{ \bar u}^4}+\frac{1}{6 \rho_{q}^4}+\frac{1}{\rho_e^4} $
\\ \hline \Trule\Brule
$\chi_2$ & $\frac{1}{ \rho_{l}^4}+\frac{3}{ \rho_q^4} $
\\ \hline \Trule\Brule
$\chi_3$ & $\frac{1}{\rho_{\bar d }^4}+\frac{1}{\rho_{\bar u}^4} +\frac{2}{\rho_q^4} $
\\ \hline \Trule\Brule
$\chi$ & $\frac{1}{\rho_{\bar d}^4}-\frac{1}{\rho_{l}^4} -\frac{2}{\rho_{\bar u}^4}+\frac{1}{\rho_q^4}+\frac{1}{\rho_e^4}$\\
\hline
\end{tabular}
\end{center}
\caption{\small The coefficients $\chi$ arise from a sum over the SM charges of $X$ particles running in the loop. The $\chi_i$, $i=1,2,3$, arise from the diagram in figure~\ref{loops}(b) with two
external SM gauge bosons of group $i$, while $\chi$ is determined by the
diagram~\ref{loops}(c) with a single hypercharge-boson on an external line.
Here $\rho_r=m_r/M$. }
\label{xi}
\end{table}

 We will not consider operators of dimension 8 containing both gauge-boson fields and the Higgs field, as well as  dimension-ten or higher operators, since their effects are  suppressed by extra powers of $M$ and/or $y$.

 The interactions in the effective action then allow the v-glueballs that cannot decay within the v-sector to decay to final states with SM particles and at most one v-glueball. This is similar to the weak decays of hadrons, such as $\pi^+\to  l\nu$, $n\to p  \nu_e e^-$, and so forth, where the  Fermi effective interaction allows otherwise stable hadrons to decay into leptons. To compute these decays, we
will only need the following factorized matrix elements\footnote{As mentioned in the introduction,  decays with no SM particles in the final state are forbidden because of kinematics and/or conserved quantum numbers.}: 
\be
\label{MatEl1}
 \ME{SM}{{{\cal O}_s^{(D-d)}}}{0} \ME{0}{{{\cal O}_v^{(d)}}}{\Theta_\kappa} \ ,
\ee
\be
\label{MatEl2}
 \ME{SM}{{{\cal O}_{s}^{(D-d)}}}{0} \ME{\Theta_{\kappa'}}{{{\cal O}_v^{(d)}}}{\Theta_\kappa} .
 \ee
   Here $d$ is the mass dimension of the operator in the v-sector,
$\langle SM|$ schematically represents a state built from Standard
Model particles, and $|\Theta_\kappa\rangle$ and
$|\Theta_{\kappa'}\rangle$ refer to v-glueball states with quantum
numbers $\kappa$, which include spin $J$, parity $P$ and
charge-conjugation $C$.   The SM part $\ME{SM}{{{\cal
O}_s^{(D-d)}}}{0}$ can be evaluated by the usual perturbative methods
of quantum field theory, but a computation of the hidden-sector matrix
elements $\ME{0}{{{\cal O}_v^{(d)}}}{\Theta_\kappa}$ and
$\ME{\Theta_{\kappa'}}{{{\cal O}_v^{(d)}}}{\Theta_\kappa}$ requires
the use of non-perturbative methods.

  \subsection{Matrix elements}

We wish to classify the non-vanishing v-sector matrix elements of the scalar operator $S\equiv \tr {\cal F}_{\mu\nu}{\cal F}^{\mu\nu}$ in (\ref{efflag}). As we saw,
the  matrix elements relevant to v-glueball transitions are given by $ \ME{0}{S}{\Theta_\kappa} $ and $  \ME{\Theta_{\kappa'}}{S }{\Theta_\kappa} $, where $\Theta_\kappa$ and $\Theta_{\kappa'}$ refer to v-glueball states with given quantum numbers.

It is convenient to write the most general possible matrix elements in terms
of a few Lorentz invariant amplitudes or form factors. The decomposition of $S$ into irreducible representations  of the Lorentz group contains only the $0^{++}$ quantum numbers~\cite{Jaffe}.  This allows the  $0^{++}$ state to decay directly  to standard model particles. For the annihilation matrix element of the $0^{++}$ v-glueball, we will write
\be	
 \,\langle
0| S|0^{++}\rangle \equiv { \bf F_{0^+}^S} 
\ee
where ${\bf F_{0^+}^S}$ is the $0^{++}$ decay constant.

 Likewise, we can  consider the transition matrix elements of the  operator $S$  between  a spin $J$ state with momentum $p$ and a spin $J'$ state with momentum $q$. For the moment we make no assumptions about the parity or charge conjugation quantum numbers of the states. Of course, if we impose parity some of the transitions may be forbidden. The matrix elements can be compactly written as
 \be
 \langle
J'|S|J\rangle\, \equiv \,\sum_i {\cal M}^{(i)}_{JJ'} \,{\bf M^{S(i)}_{J J'}} 
\ee
 where now ${\bf M^{S(i)}_{J J'}}$ is the transition matrix, which depends on the transferred momentum, and ${\cal M}^{(i)}_{JJ'} $ is determined by the Lorentz representations of $|J\rangle$ and $|J'\rangle$. In Appendix C we have listed  ${\cal M}^{(i)}_{JJ'}$ for the simplest cases considered later in this work.

The main uncertainties in the study of decays of the v-glueballs stem from the evaluation of the transition matrix elements ${\bf M^{S(i)}_{J J'}}$ which at present are unknown. Many of these could be in principle be determined by additional lattice computations. In spite of these uncertainties, we will still be able to obtain
many interesting and robust results.

\section{Decay rates}
\setcounter{equation}{0}
An interesting feature of the model described above 
is that it has a spectrum of v-glueballs  which can interact very weakly 
with the particles in the standard model through effective operators of different mass dimensions. The dimension-six operator, as we see below,
permits the v-glueballs to decay directly into standard model particles (the $0^{++}$) or  radiatively by emitting a Higgs boson (the  $2^{++}$, $2^{-+}$, $3^{++}$, $3^{+-}$, $2^{+-}$, $0^{+-}$, $1^{--}$, $2^{--}$, $3^{--}$).  On the other hand, dimension-eight operators contribute with additional decay modes. As shown in ~\cite{Juknevich:2009ji}, these include direct annihilations into SM gauge-boson pairs (the $0^{++}$, $2^{++}$, $0^{-+}$ and $2^{-+}$) 
or  $C$-changing  radiative transitions with emission of a photon or a $Z$ boson (all others).

In this section, we will compute the decay rates for most of the v-glueballs in figure \ref{spectrum}.
 We  will first present our results for the v-glueball decays induced by dimension-six operators given in (\ref{efflag}). Then we will make a quick review of the results of ~\cite{Juknevich:2009ji} for dimension-eight operators that  are relevant for our work. 

\subsection{V-glueball decays by dimension-six operators}
We begin with the $0^{++}$ v-glueball, which can be created  by the  operator $S$. The  dimension-six coupling  leads to  $0^{++}\to \zeta\zeta$ annihilations, where $\zeta$ denote any of the final states of the Higgs boson. Above the threshold for Higgs boson pair production, the  $0^{++}\to hh$ decay can also proceed with a sizeable rate. Other decay modes induced by dimension-six couplings include   processes of the form $\Theta_\kappa\to \Theta_{\kappa'} h$, where $\Theta_\kappa$, $\Theta_{\kappa'}$ denote two v-glueballs with given quantum numbers.  As an example of a Higgs-radiative decay,  we will present the computation of the $2^{++}\to 0^{++} h$ decay with some detail. Then we will consider the general case  $\Theta_\kappa\to \Theta_{\kappa'} h$ for v-glueballs with arbitrary quantum numbers. In the end, we make some comments on the $0^{-+}$ and $1^{+-}$ v-glueballs, which are the only ones that are not permitted to decay via dimension-six operators.

\subsubsection*{Annihilation of the $0^{++}$ state}
The scalar state can be created or destroyed by the $S$ operator. Then, 
the effective interaction (\ref{efflag}) allows the decay of the $0^{++}$ state via $s$-channel Higgs-boson exchange $0^{++}\rightarrow h^{*}\to \zeta \zeta$, where 
$\zeta$  collectively denotes  a standard model particle. 
According to (\ref{efflag}), the amplitude for this decay reads
\be
\frac{y^2  \, \alpha_v\, }{3\pi M^2 } \,\langle
\zeta \zeta|   m_f  \bar f f+m_Z^2  Z_\mu Z^\mu + 2m_W^2  W_\mu^+ W^{\mu-}|\, 0 \rangle \,\frac{1}{m_{H}^2-m_0^2}\, \langle
0| S|0^{++}\rangle 
\ee
where $\alpha_v = g_v^2/(4\pi)$ and $m_H$ is the Higgs mass.
The width of the decay is given by
\be \label{wid0}
\Gamma_{0^{++}\rightarrow \zeta \zeta} = \left(\frac{y^2 \,v_H \,\alpha_v \,{\bf F_{0^+}^S}}{3 \pi M^2 (m_H^2-m_0^2)}   \right)^2 \Gamma^{SM}_{h \rightarrow \zeta \zeta}(m_{0^+}^2)
\ee
where ${\bf F_{0^+}^S} \equiv  \,\langle
0|S|0^{++}\rangle$ is the $0^{++}$ decay constant.  Here $\Gamma^{SM}_{h \rightarrow \zeta \zeta}(m_{0^+}^2)$ is the width for the decay
$h \rightarrow \zeta\zeta$ for a standard model Higgs boson with a mass $m_{0^+}$.
Then, formula (\ref{wid0})  implies that the branching ratios of the  $0^{++}$ are  those of the SM Higgs boson in the range of mass of interest.  Expressions for the branching ratios  and full width of the Higgs boson can be found in the literature, for masses ranging from a few $\MeV$ up to $1\TeV$ (for a review, see \cite{hunter}). 

Although these are standard model tree-level results, we should also remark that the range of validity of (\ref{wid0}) is beyond the simple perturbative QCD  domain. For masses below $2-3 \GeV$, the $0^{++}$ v-glueball can decay  into a pair of hadrons via its interaction with two gluons through a top-quark loop or its interaction with quarks.  The hadronization of these quarks and gluons is a rather complex and non-perturbative process. However, the branching ratios are still given by (\ref{wid0}), with $\zeta$ now running over the possible hadrons in the final states, such as $\pi$, $K$, and so forth.

If $m_{0^+}> 2 m_h$,  the decay channel $0^{++}\to hh$  opens up and the partial width is given by
\be \label{wid0hh}
\Gamma_{0^{+}\rightarrow hh} =\frac{1}{32 \pi  m_{0^+}} \left(\frac{y^2\,\alpha_v \, {\bf F_{0^+}^S}}{ 3\pi\,M^2}\right)^2 \left(1+\frac{3 m_H^2}{(m_{0^+}^2-m_H^2)}\right)^2 \left[g(m_H^2,m_H^2;m_{0^+}^2) \right]^{1/2}
\ee
where $g(x,y;z)\equiv  \left(1-x/z-y/z\right)^2-4 xy/z^2$.
The off-shell decays $0^{++}\to h^* h $ and $0^+\to h^*h^*$ are also possible in the intermediate mass range. However, these channels receive an extra suppression from the smaller available phase space, so it is reasonable to expect they will have little effect on the full $0^{++}$ width.

As $m_{0^+}$ becomes larger than $2 m_t$ and $2 m_H$, we have the following approximate relationship among the dominant decay rates
\be
\Gamma_{W^+ W^-}:\Gamma_{ZZ}:\Gamma_{hh}:\Gamma_{t\bar t}=2:1:1:3 \,x_t \, \sqrt{1-x_t}
\ee
where $x_t=4 m_t^2/m_{0^+}^2$.

\subsubsection*{An example of  Higgs-radiative decays: the $2^{++}\to 0^{++} h$ case}

As an illustration of the computation of the decay rates for transitions with the emission of Higgs boson,  let us first consider  in some detail the decay of the $2^{++}$ v-glueball. Contrarily to the $0^{++}$, direct annihilation of the $2^{++}$ v-glueball into Higgs bosons would require an operator in the effective action of dimension $D=8$, and is hence negligible.
Instead,  for $m_{2^+}- m_{0^+}> m_H$, the $S$ operator induces the $2^{++}\rightarrow 0^{++} h$ decay. In a parton-model picture, one may imagine this process as being caused by the radiative emission of a Higgs boson through $g_v\to g_v h$, with one spectator v-gluon going to the final state.
 The amplitude of this two-body  decay is given by
\be
\frac{y^2\, \alpha_v }{3 \pi M^2}\ME {h} { h} {0} \ME{0^{++}}{S}{2^{++}} = \frac{y^2 }{3 \pi M^2}\, {\bf M_{0^+2^+}^S}\, \epsilon_{\mu\nu} q^\mu q^\nu
\ee
where $q^{\mu}$ is the momentum of the $0^{++}$ and $\epsilon_{\mu\nu}$ the polarization of $2^{++}$. Here ${\bf M}_{0^+2^+}^S$ denotes the transition matrix element defined by ${\bf M}_{0^+2^+}^S \epsilon_{\mu\nu} q^\mu q^\nu = \ME{0^{++}}{S}{2^{++}}$; for simplicity  we have assumed  it is independent of the transferred momentum.  The corresponding width of the decay reads
\be \label{2dtensor}
\Gamma = \frac{y^4 \,\alpha_v^2 }{480 \pi (3\pi)^2 M^4} m_{2^+}^3 \left[g (m_{0^+}^2,m_H^2 ;m_{2^+}^2)\right]^{5/2} ({\bf M_{0^+2^+}^S})^2.
\ee

 If the  $2^{++}$ is not heavy enough, $m_{2^+}- m_{0^+}< m_H$, it can decay to the $0^{++}$ and SM particles through the emission of an off-shell Higgs. The corresponding amplitude for the three-body decay $2^{++}\to 0^{++} \zeta\zeta$ is
\be
\frac{y^2  \alpha_v}{3\pi M^2}\,\langle
\zeta\zeta|    m_f  \bar f f+m_Z^2  Z_\mu Z^\mu + 2m_W^2  W_\mu^+ W^{\mu-} |\, 0 \rangle \, \frac{1}{k^2-m_H^2+i \Gamma_h^{SM} m_H}\, \langle
0^{++}| S|2^{++}\rangle
\ee
where $k^2$ is the  transferred momentum. The width  is given by
\be\label{3btesor}
\Gamma_{2^{++}\to 0^{++}\zeta\zeta}=\frac{m_{2^+}^3 ({\bf M_{0^+2^+}^S})^2}{160\pi^2}\left(\frac{y^2 \, \alpha_v v_H}{3\pi M^2 }\right)^2  \int dm_{12}^2 \frac{\left[g(m_{0^+}^2,m_{12}^2;m_{2^+}^2)\right]^{5/2}\Gamma_{h\to \zeta\zeta}^{SM}(m_{12})}{(m_{12}^2-m_H^2)^2 +(\Gamma_h^{SM})^2 \,m_H^2}
\ee
where the limits of integration are set by available phase space.

\subsubsection*{Higgs-radiative decays. General case $J\to J' h$}

We now consider the 2-body decays $\Theta_J\to \Theta_{J'}\,h$, in which $\Theta_J$, $\Theta_{J'}$ are v-glueballs with spin $J$, $J'$ respectively. For the moment, we make no reference to the parity of the v-glueballs and proceed generally. Parity conservation will force some of the matrix elements to be zero. Making use of the general formulas for the matrix elements (\ref{m32})-(\ref{m10}), we arrive at the decay rate

\be\label{jjh}  
\Gamma_{J\to J' h }^{(i)}(m_H^2)  = \frac{ y^4  \,v_h^2\,\alpha_v^2}{16 \pi (3\pi)^2 M^4 \,m_J\,(2 J+1)} |{\bf M_{J,J'}^S}|^2 \, \Gamma_{JJ'}^{(i)} \, \left[g(m_{J'}^2,m_H^2;m_J^2)\right]^{1/2}
\ee
where $i$ runs over the various form factors. The coefficients $\Gamma_{JJ'}^{(i)}$  are dimensionless functions of the masses and depend on the angular momentum transfer associated with each transition. They are summarized in table \ref{tensors2}.

\begin{table}[ht]
\begin{center}
\begin{tabular}[c]{|c|c|c|c|c|c|}\hline \Trule\Brule 
$i$&1 &2&3&4&5 
\\ \hline\Trule\Brule
${\Gamma}^{(i)}_{32}$ &$\frac{x}{15}(4x^2+28x+35)$&$\frac{4 x^3}{15}(x+2)$&$\frac{4 x^5}{15}$&$\frac{2 x^2}{15}(2 x +7)$&$\frac{4 x^4}{15}$\\ \hline\Trule\Brule
${\Gamma}^{(i)}_{31}$ &$\frac{2 x^2}{15}(3x+7)$&$\frac{2 x^4}{5}$&$\frac{8 x^3}{15}$&-&-\\ \hline\Trule\Brule
${\Gamma}^{(i)}_{30}$ &$\frac{2 x^3}{5}$&-&-&-&- \\ \hline\Trule\Brule
${\Gamma}^{(i)}_{22}$ &$\frac{1}{9}(4 x^2+ 30 x +45) $&$\frac{x^2}{18}(8 x + 17)$&$\frac{4 x^4}{9}$&$\frac{x}{2}(x+5)$&$\frac{x^3}{2}$\\ \hline\Trule\Brule
${\Gamma}^{(i)}_{21}$ &$\frac{x}{3}(2 x+5)$&$\frac{2 x^3}{3}$&$x^2$&-&-\\ \hline\Trule\Brule
${\Gamma}^{(i)}_{20}$ &$\frac{2 x^2}{3}$&-&-&-&- \\ \hline\Trule\Brule
${\Gamma}^{(i)}_{11}$ &$x+3$&$x^2$&$2x$&-&-\\ \hline\Trule\Brule
${\Gamma}^{(i)}_{10}$ &$x$&-&-&-&- \\ \hline
\end{tabular}
\end{center}
\caption{The coefficients $\Gamma_{JJ'}^{(i)}$ arise from the average of the   squared  matrix elements. We denote $x= \frac{m_J^2}{4 m_{J'}^2} g(m_{J'}^2,m_H^2 ;m_J^2)$. The dashes denote those cases where form factors are absent.}
\label{tensors2}
\end{table}
Below the threshold for Higgs boson production, the decay rate for the 3-body decay $\Theta_J\to \Theta_{J'}\,\zeta\zeta$ reads,
\be\label{3body}
\Gamma_{J\to J' \zeta\zeta}= \frac{1}{\pi } \int dm_{12}^2 m_{12} \Gamma_{J\to J' h}^{(i)}(m_{12}) \frac{1}{\Delta(m_{12}^2,m_H^2)} \Gamma_{h\to \zeta\zeta}^{SM}(m_{12}).
\ee
where $\Delta(m_{12}^2,m_H^2) = (m_{12}^2-m_H^2)^2+ m_H^2 (\Gamma^{SM}_h)^2$. The integration automatically includes the case where the radiated Higgs boson can be close to  onshell.

Some of the transitions may be parity-forbidden (e.g. $0^{-+}\to 1^{+-} h$). However, it is a straightforward exercise to check that for each v-glueball $\Theta_\kappa$ in figure \ref{spectrum} , there exists at least one other v-glueball $\Theta_{\kappa'}$ such that the transition $\Theta_\kappa\to \Theta_{\kappa'}h $ is allowed,
with a rate which is given by (\ref{jjh})-(\ref{3body}). The only exceptions are the $0^{+-}$ and $1^{+-}$ v-glueballs  that we discuss next.

\subsubsection*{Decays of the pseudoscalar and pseudovector}
The only states that are not allowed to decay via dimension-six operators are the $1^{+-}$ and the $0^{-+}$ v-glueballs.

Since the $1^{+-}$ state is the lightest state in the $C$-odd sector, it  necessarily has to decay  to a v-glueball of opposite  $C$. One possibility is that the $1^{+-}$ decays by radiatively emitting a Higgs boson (e.g. $1^{+-}\to 0^{-+} h$). However, this decay mode would violate $C$ and, hence, is forbidden.  

A second way to induce decays of the $1^{+-}$ v-glueball would be
via its coupling to the hypercharge current $H^\dagger D_\mu H$. In this case, there are three v-sector operators that can be contracted with $H^\dagger D_\mu H$, namely, 
\be\label{vectorops}
\tr {\cal F}_{\alpha\beta} D_\mu {\cal F}^{\alpha\beta},\;\;\;\;\tr {\cal F}^{\alpha\beta} D_\beta {\cal F}_{\alpha\mu},\;\;\;\;\tr {\cal F}_{\mu\beta} D_\alpha {\cal F}^{\alpha\beta}.
\ee
However, one can see that these operators cannot induce $C$-changing transitions.
 First, notice that classically  $\tr {\cal F}_{\alpha\beta} D_\mu {\cal F}^{\alpha\beta}=\frac12  \partial_\mu \tr {\cal F}_{\alpha\beta}  {\cal F}^{\alpha\beta}$. As explained in \cite{Jaffe}, this implies that the transitions  induced by the first operator in (\ref{vectorops}) are not new. They are just the same ones created by the  operator $S$.
 Likewise,  using  equations of motion and   conservation of the energy-momentum tensor,
the second operator in (\ref{vectorops}) can be related to 
a total derivative of the operator $S$.
Finally,  equations of motion also imply that the last operator in (\ref{vectorops}) vanishes identically. Therefore, up to operators of total mass dimension six, the $1^{+-}$ state is stable. 
In the next subsection, we shall see that dimension-eight operators can induce photon-radiative decays of the $1^{+-}$ v-glueball to $C$-even v-glueballs.

On the other hand, $C$-invariance by itself would allow the Higgs-radiative transition $0^{-+}\to 0^{++} h$. However, this decay could not conserve both angular momentum and parity $P$: since the initial state is $0^{-}$, angular momentum conservation requires the orbital angular momentum of the $0^+$ and $h$ final state to be $L=0$, which in turn requires total parity $P=+1$, rather than $P=-1$ as demanded by parity conservation. This decay mode is thus forbidden. This is analogous to the way that the $\eta$ meson strong interaction mode $\eta\to\pi\pi$ is forbidden in the SM. A similar argument shows that the three-body decay $0^{-+}\to 0^{++} hh$ is not permitted. In this case, the corresponding $\eta\to\pi\pi\pi$ decay is allowed in the SM because, contrary to the $0^{++}$ and $h$, the pions have intrinsic parity $-1$. As we will see in section 6, this line of argument alters if we relax our assumptions and allow for $P$-violating couplings.

\subsection{V-glueball decays by dimension-eight operators}
The operators in the effective action (\ref{eff_lag full2}) induce decays for all of the v-glueball states in figure \ref{spectrum}. 
 The  lightest states in the C-even sector (the $0^{++}$, $2^{++}$, $0^{-+}$ and $2^{-+}$)  can directly decay to pairs of standard model gauge bosons ($gg$, $\gamma\gamma$, $ZZ$, $WW$ or $\gamma Z$) via the $S$, $P$, $T$ and $L$ operators.
All other states can decay by radiatively  emitting  a photon or, to a lesser extent, a $Z$ boson, via the $d=6$ $D=8$ operators $\Omega_{\mu\nu}$. 
Here we briefly summarize the results of ~\cite{Juknevich:2009ji} concerning the computation of the decay rates induced by $D=8$ operators.

In order to retain simplicity, we will often assume, in the subsequent discussion,
the $X$ fields form approximately degenerate multiplets of $SU(5)$, i.e. $\rho_r\approx 1$. 
When there is a large hierarchy between the colored and the uncolored $X$ particles, $\rho_{\bar d}\gg \rho_l=\rho_e$ and $\rho_{\bar u}\approx \rho_q\gg\rho_e$, the decay pattern becomes slightly more complicated because the decay channels into $\gamma\gamma$, $\gamma Z$, $ZZ$ and $WW$ may all play a role and even dominate in some regions of the parameter space.  We will make some comments later in this paper on some of the interesting phenomenology that arises in this regime.
 
For the decay of the $0^{++}$, $2^{++}$, $0^{-+}$ and $2^{-+}$ v-glueballs into gluons we have the following rates,
 \be\label{widgg}
\Gamma({0^{+}\rightarrow gg}) = \frac{\alpha_s^2\alpha_v^2}{2\pi
M^8}\,\chi_3^2  \left(\frac{1}{60}\right)^2 m_{0^+}^3({\bf F_{0^{++}}^S})^2.
\ee
\be
\Gamma({2^{+}\rightarrow gg}) = \frac{\alpha_s^2\alpha_v^2}{20\pi
M^8}\,\chi_3^2m_{2^+}^3\left[\frac{1}{2}\, \left(\frac{11}{45}\right)^2 ({\bf
F_{2^{++}}^T})^2 +\frac{4}{3}\,\left(\frac{1}{30}\right)^2 ({\bf
F_{2^{++}}^L})^2\right] ,
\ee
\be
\label{p2gg}
\Gamma({0^{-}\rightarrow gg}) = \frac{\alpha_s^2\alpha_v^2}{2\pi
M^8}\,\chi_3^2 \left(\frac{2}{45}\right)^2m_{0^-}^3({\bf F_{0^{-+}}^P})^2
\ee
\be
\label{l2gg}
\Gamma({2^{-}\to gg}) = \frac{\alpha_s^2\alpha_v^2}{15\pi
M^8}\,{\chi_3}^2m_{2^-}^3\left(\frac{1}{30}\right)^2  ({\bf
F_{2^{-+}}^L})^2
\ee
where $\alpha_s=g_3^2/(4\pi)$ is the QCD coupling constant and the coefficient $\chi_3$ is given in table \ref{xi}. Of great interest is the branching fraction into two photons, 
\be \label{braa}
\frac{\Gamma(\Theta\to \gamma\gamma)}{\Gamma(\Theta\to gg)} = \frac12\frac{\alpha^2}{\alpha_s^2}\frac{\chi_{\gamma}^2}{\chi_3^2}
\ee 
for $\Theta= 0^{++}, 2^{++}, 0^{-+}, 2^{-+}$. Here $\alpha$ is the fine structure constant and $\chi_\gamma\equiv \chi_1+\chi_2/2$, where  $\chi_1$ and $\chi_2$ are shown in table \ref{xi}. The expressions for the branching fractions into electroweak bosons are  omitted for the sake of brevity but can be found in~\cite{Juknevich:2009ji}. Some comments on the weak boson decay modes will be given in section 5.

In the C-odd sector, all the v-glueballs decay radiatively with the emission of a photon to the lightest v-glueballs in the C-even sector. Direct annihilations into three SM gauge bosons are suppressed since they would be induced by dimension-twelve operators. One can also show that three-body decays are suppressed due to the small  phase space that is available for these decays~\cite{Juknevich:2009ji}.

The lightest states in the C-odd sector are the pseudovector $1^{+-}$ and the vector $1^{--}$. 
The width of the decay $1^{+-}\rightarrow 0^{++}+ \gamma$ is given by
\be
\label{h2sg}
\Gamma_{1^{+}\to 0^{+}+\gamma}= \frac{\alpha\alpha_{v}^3}{24\pi
M^8}\,{\chi^2}\,\frac{(m_{1^+}^2-m_{0^+}^2)^3}{
m_{1^+}^3}\,({\bf M_{1^{+-}0^{++}}^\Omega})^2.
\ee
A similar expression holds for the decay $1^{--}\rightarrow 0^{++}+ \gamma$ 
\be
\label{1--tog}
\Gamma_{1^{-}\to 0^{+} + \gamma} = \frac{\alpha\alpha_{v}^3}{24\pi
M^8}\,{\chi^2}\,\frac{(m_{1^-}^2-m_{0^+}^2)^3}{
m_{1^-}^3}\,({\bf M_{1^{--}0^{++}}^\Omega})^2.
\ee

For the $1^{--}$ state, annihilation to SM fermions via an off-shell $\gamma$ or $Z$ is also possible, with branching ratio
\be
\frac{\Gamma_{1^{-}\to\gamma^*/Z^*\to
f\bar{f}}}{\Gamma_{1^{-}\rightarrow 0^{+}+\gamma}} = \frac{16 \pi \alpha}{\cos^4\theta_W}\,(Y_L^2+Y_R^2)
\left(\frac{m_{1^-}^2}{m_{1^-}^2-m_{0^+}^2}\right)^3\left(\frac{{\bf
F_{1^{--}}^\Omega}}{{\bf M_{1^{--}0^{++}}^\Omega}}\right)^2.
\ee
Here $Y_L$ and $Y_R$ are left and right hypercharge of the emitted fermions. Decay to electrons and muons will be reconstructable as a resonance, so despite its uncertain branching fractions, this decay mode is mode is worthy of careful consideration.

One can also generalize formulas (\ref{h2sg}) and (\ref{1--tog}) to include the radiative decays of the heavier C-odd v-glueballs. Easy computations show that
\be
\Gamma_{J\to 0^{++} \gamma} =\frac{\alpha\,\alpha_v^3\,\chi^2}{4\pi\, M^8} \frac{(J+1)\,(J!)^2}{2^J \,J\,(2J)!\,(2J+1)} \frac{(m_J^2-m_{0^+}^2)^{2J+1}}{m_J^{2J+1} m_{0^+}^{2J-2}} \,({\bf M_{J 0^{++}}^\Omega})^2
\ee
\begin{multline}
\Gamma_{J\to 2^{++} \gamma}\, =\,\frac{\alpha\alpha_v^3}{48\pi\,M^8} \,\chi^2 \,\frac{(m_J^2-m_{2^+}^2)^{2J+1}} {m_J^{2J+3} m_{2^+}^{2J}}\, \frac{2^{J-7} (J!)^2 } {3 \,J \, (2 J)!\,(2J+1)} \,\\\left(2 (71 J+65) m_J^2 m_{2^{++}}^2+3 (5 J+3) m_J^4+3 (5 J+3) m_{2^{++}}^J\right) \,({\bf M_{J 2^{++}}^\Omega})^2.
\end{multline}
with similar expressions  for the modes $J\to 0^{-+} \gamma$ and $J\to 2^{-+}\gamma$.

\subsection{Summary of decays}
In table \ref{decay-modes} we summarize  the  final states for the most important decay channels of the  v-glueballs  in figure \ref{spectrum} for $D=6$ and $D=8$ operators. 
\begin{table}[ht]
\begin{center}
\begin{tabular}[c]{|c|c|c|c|}\hline \Trule\Brule 
State & $D=6$ operators &$D=8$ operators
\\ \hline\Trule\Brule
$0^{++}$&$bb$, $W^+W^-$, $ZZ$, $hh$&$gg$, $WW$, $ZZ$, $Z\gamma$, $\gamma\gamma$  \\ \hline\Trule\Brule
$2^{\pm+}$ &$0^{\pm+}h(h^*)$&$gg$,  $WW$, $ZZ$, $Z\gamma$, $\gamma\gamma$ \\ \hline\Trule\Brule
$0^{-+}$& -&$gg$,  $WW$, $ZZ$, $Z\gamma$, $\gamma\gamma$\\ \hline\Trule\Brule

$3^{++}$&$0^{-+}h$, $2^{\pm+} h(h^*)$& $0^{-+}gg$ $ 2^{++}gg$, $1^{+-}\gamma$  \\ \hline\Trule\Brule
$1^{+-}$& - &$ 0^{\pm+}\gamma$, $2^{-+}\gamma$ \\  \hline\Trule\Brule

$1^{--}$&$1^{+-}h(h^*)$&$ 0^{\pm+}\gamma$, $2^{\pm+}\gamma$, $ff$ \\ \hline\Trule\Brule
$0^{+-}$, $2^{+-}$, $3^{+-}$&$J^{P-} h(h^*)$&$ 0^{\pm+}\gamma$, $2^{\pm+}\gamma$\\ 
$2^{--}$, $3^{--}$&&\\  \hline
\end{tabular}
\end{center}
\caption{Possible final states of the various  v-glueballs in figure \ref{spectrum} generated by  $D=6$ and $D=8$ operators. Note the absence of Higgs-mediated decay modes for the $0^{-+}$ and $1^{+-}$ v-glueballs. Here $J^{P-}$ denotes a $C$-odd v-glueball state.}
\label{decay-modes}
\end{table}
   We therefore see the presence of operators of different mass dimensions opens a plethora of decay modes, which is particularly interesting from the phenomenological point of view, but complex to analyze. It is the purpose of the next two sections to disentangle the effects from $D=6$ and $D=8$ operators, and extract the most frequent decay modes. 

\section{Constraints on new physics}
\setcounter{equation}{0}
In this section, we discuss direct experimental constraints  on the operators used in this paper from Tevatron searches for new physics, as well as potential limits from precision electroweak measurements. 
\subsection{Direct searches}

In order to translate nonobservation of $X$ particle events above the standard model background   into bounds on the $X$ mass, it is 
 important to understand the  possible collider signals associated with the production of $X$ particles at hadron colliders such as the Tevatron or the LHC. 
 As was pointed out in \cite{Okun,quinn}, and more recently in \cite{SZ,KLN}, theories with quirks like the  $X$ particles give rise to a quite unusual dynamics. 
 Here we briefly rephrase and elaborate some of the results of  \cite{SZ,KLN} to agree with the regime we are considering in this paper.

 At the Tevatron, color-singlet  $X$ particles could be pair produced through an off-shell photon,  $Z$, or $W$. If the  $X$ particles are colored, their pair production through an off-shell gluon is also possible.
 To a great extent, the production rate is only dependent on the gauge couplings of the $X$ particles and so is fixed by their standard model quantum numbers.
 The $X$ and $\bar X$ are initially produced with considerable kinetic energy.
 Due to v-confinement, they form  a bound state of $X$-onium with an $SU(n_v)$ string (flux tube) being stretched between the $X$ and the $\bar X$. As the $X$ and $\bar X$ move apart, the potential energy stored in  the string increases, but contrary to QCD strings, the $SU(n_v)$ string cannot break, since there are no light v-quarks in the v-sector whereby the string can split. Instead,   the $X$ and $\bar X$  oscillate back and forth losing their kinetic energy and angular momentum  by radiating some combination of photons, hadrons and  v-glueballs. Eventually the $X$ and $\bar X$ recombine and annihilate back into lighter states. 

In the case of colored $X$ particles, the annihilation is most often into v-gluons, which at long-distance become two or more v-glueballs. As explained in section 3, the v-glueballs can then decay via a loop of $X$ particles, producing photons, gluons, bottom quarks, and so forth. As will be seen in section 5, v-glueball decays typically result in an observable signal, providing $\Lambda_v\gtrsim 1\GeV$.
For color-singlet $X$ particles that are produced via an off-shell photon or $Z$, the primary annihilation channel is also to v-glueballs, which can decay back to SM particles.  However, the dominant cross-section for production of color-singlet  $X$ particles   is through an off-shell $W$ boson~\cite{FoldedSUSY}. The resulting $X^{+}\bar X^0$ bound state is electrically charged and cannot annihilate into v-glueballs alone\footnote{One can show that the splitting between $X^+$ and $X^0$ is so small that $X^+$ is stable against decay to $X^0$ on the time scales of interest; for example, see~\cite{FoldedSUSY}. }. Instead, it may annihilate to leptons or quarks via an off-shell $W$, to $W\gamma$ or $WZ$, or to $W g_v g_v$, that is $W$ plus some v-glueballs. Other contributions to the final states  arise from the v-glueballs radiated during the $X$-onium relaxation~\cite{SZ,FoldedSUSY,KLN}. These  might help increase the discovery reach and deserve further investigation which we leave for future work.

The $W\gamma$ process might allow $X$-onium to be discovered directly. Events with a photon, lepton and missing energy ( $l+\gamma+\MET$) would reveal a $X$-onium resonance in the transverse mass distribution above the $W\gamma$ background. On the other hand, when the v-glueballs decay, they may produce spectacular collider signals, such as $p\bar p \to X\bar X\to \gamma\gamma\gamma\gamma,\gamma\gamma\tau^+\tau^-$. The $4\gamma$ channel, in particular, is very clean, since it is essentially background free, and would reveal the v-glueballs, if the rate is large enough to be detected.

There have not been specific searches for quirks as yet. Present limits are in general very mild because the cross-section for production of quirks at the Tevatron becomes very small once the quirk masses are larger than $\sim 250\GeV$ and because not all the collected data has been analyzed yet. Below we present bounds on the masses of the $X$ particles  from CDF searches for new physics in $ l +\gamma+\MET$ events and in searches for anomalous production of two photons and at least one more energetic, isolated and well-identified object ($\tau$ lepton or $\gamma$). 
This analysis by no means is exhaustive, but is meant to illustrate potential discovery signals and the bounds that can arise.

Our estimates below are based on the following assumptions. First, we determine the rate for production of $X$ particles through perturbative computations; the cross-section $\sigma_{X\bar X}$ is estimated as a function of the mass $M_X$ and for $n_v=2$, both for the cases of colored and uncolored particles. Second, motivated by  a thermal model of fragmentation~\cite{Falkowski:2009yz}, we have estimated the relative probability of producing any v-glueball in $X$-onium annihilations assuming production of the $0^{++}$ at $50-60\%$, and that of the $2^{++}$, $0^{-+}$ and  $2^{-+}$ totaling  $40-50\%$ in all. Finally, for the purpose of obtaining an approximate bound we assume a $100\%$ efficiency for detecting v-glueballs. The last two assumptions are unlikely to be accurate, as we will discuss more thoroughly in our LHC study, but they are intended to give a rough estimate for the bounds. 

\vspace{2mm}
(i) $l+\gamma+\MET$:
The CDF collaboration has searched for the anomalous production of events containing a high-transverse momentum charged lepton and photon, accompanied by missing transverse energy~\cite{cdf1}.  Using  $929 {\rm pb}^{-1}$ of CDF Run II data, these searches exclude  $X$ masses below $200\GeV$ for color-singlet $X$ particles and $250\GeV$ for colored $X$ particles, under the assumption that the $X$-onium has lost most of its energy by radiation of photons and/or v-glueballs and the annihilation takes place at or near the ground state.

 \vspace{2mm}
(ii) $2\gamma+ \gamma$ and $2\gamma+ \tau$:
 The CDF collaboration has also searched for the inclusive production of diphoton events~\cite{cdf2}.  Using $1155 {\,\rm pb}^{-1}$ of integrated luminosity, the diphoton plus third photon search places lower limits on the $X$ mass below $200\GeV$ for color-singlet $X$ particles and $y\sim 0$. The corresponding  limit on the $X$ mass  for  colored $X$ particles turns out to be less severe, roughly below $150 \GeV$,  because,  in this case, the branching fraction to photons is $\lesssim 0.35 \%$. Similarly, using $2014 {\,\rm pb}^{-1}$, the diphoton plus tau search places lower limits on the $X$ mass for  $y\sim 1$ below $250\GeV$ for color-singlet $X$ particles and $175\GeV$ for colored particles.

\vspace{2mm}

Some comments are in order. The mass bounds from potential v-glueball signals are necessarily model dependent.  Unfortunately, unlike our computations of the branching ratios, there is no way to make a reliable estimate of the probability for producing any given v-glueball state in $X$-onium annihilation, since  phenomenological input from QCD is not relevant to the pure-glue case.
V-glueballs could be created during the relaxation of  $X$-oniums,  and during their annihilation, so events with more than two v-glueballs might be common.  Generally, with more than two v-glueballs produced in each $X$-onium annihilation, events with multiple photons would be common and easier to observe. However, the events may be more cluttered and the energy of the photons lower, leading potentially to lower detection efficiencies. Since we cannot model this reliably, in the simple analysis of this paper, we will rely on the model independent bounds from the production of $X$-onium through $s$-channel $W^{\pm}$ outlined above.

Moreover, the bounds from the $W+$ photon final state assume that the $X$-onium state decays to its ground state before annihilating. However, it has been argued in \cite{KLN} that  the annihilation may often occur at higher energy, before the ground state is reached. With many $X$-onium states annihilating at different energies, the signal in the transverse mass distribution will be diluted. Besides, annihilation to two fermions via an off-shell $W$ or to $W$+ v-glueballs becomes increasingly significant at higher energies. While the branching ratios to $W\gamma/Z$, $f\bar f$ and $W$+v-glueballs can be estimated, the wide range of $X$-onium states and the uncertain annihilation probability as a function of the energy make any precise evaluation almost impossible. Therefore, the Tevatron limits are in fact weaker than suggested above.

In passing, we note that the bounds on a fourth generation of particles from Higgs searches at the Tevatron do not apply to heavy fermions that get only part of their mass from electroweak symmetry breaking, such as the $X$ particles in our paper.
For an additional pair of fermions that get most of their mass from electroweak symmetry breaking, the Higgs production cross-section $\sigma_{gg\to H}$ is known to increase by a factor of roughly 9, making a dramatic effect on Higgs physics~\cite{Kribs:2007nz}. This allowed CDF and D0 to rule out fourth generation quarks for a Higgs mass in the window $145\GeV<m_H< 185\GeV$. However, for heavy particles that get only part of their mass by electroweak symmetry breaking, their contribution to the cross-section  $\sigma_{gg\to H}$ decreases and the bounds are generally much weaker or absent.

Before moving on to the discussion of electroweak precision constraints we should briefly comment on the possibility of  producing v-glueballs in Higgs decays. Up to now  we have considered the production of v-glueballs as a by-product of $X$-onium annihilation and relaxation. However, there is another possibility for this production to occur. For sufficiently small v-glueball masses, the interaction (\ref{efflag}) can also mediate processes such as $h\to \Theta_\kappa\Theta_\kappa$ and $h\to \Theta_\kappa\Theta_\kappa\Theta_\kappa$. 
Assuming only minor phase space suppression, the branching ratio to v-glueballs for a SM Higgs below $140\GeV$ and $n_v=2-4$ is of order $(\alpha_v y^2 v_h^2 m_h/3\pi M^2 m_b)^2\sim 10^{-3}-10^{-2}$. This implies a cross-section for producing v-glueballs at the Tevatron   of order a few fb. Even though the cross-section is small,  when combined with $X$-onium annihilations, the  production of v-glueballs in Higgs decays may be useful in the low mass range $m_0\simeq 1-70\GeV$.
 The produced v-glueballs can then decay to a wide variety of final states, including $b$ quarks, $\tau$ leptons, and multiple photons, among others; for small v-glueball masses, a significant fraction of these decays may occur with displaced vertices.  At the Tevatron, there may be potential to observe a few events in the $b\bar b \tau^+\tau^-$ or $b \bar b b \bar b$ channels, although with limited statistics. In the case of displaced vertices, the  D0 collaboration has searched for pair production of neutral long-lived particles decaying to a $b\bar b$ pair, and found no significant excess above the background~\cite{Abazov:2009ik}.   This search could only exclude branching fractions of order 1, so  that in our case it does not imply any bounds.
 At the LHC, the background problem is  more severe, but with a larger  cross-section and a larger integrated luminosity than at the Tevatron, the observation of a few v-glueball events may also be possible. Depending on kinematics, the case of photons that originate away from the primary interaction point may also play an interesting role.  A search for displaced photons has been done at D0~\cite{Abazov:2008zm}, and may be possible at ATLAS, due to the longitudinal structure of the electromagnetic calorimeter (ECAL). Besides, CDF and CMS allow the identification of delayed photons from long-lived particles using  ECAL timing information (see \cite{Aaltonen:2008dm} for results obtained using the CDF detector).  The question then is whether
at a rate $\sim 10^{-3}-10^{-2}$ the LHC will have some sensitivity.  Keeping  these non-standard Higgs decays in mind we leave the analysis of the discovery reach  for future work.

\subsection{Electroweak oblique corrections}

If new heavy particles exist, they  can manifest in the standard model in terms of corrections to the gauge-boson self-energies. When the new physics scale is much larger than $M_Z$, this effect can be described by just three parameters $(S,T,U)$ at the one-loop level ~\cite{Lynn:1985sq,Peskin:1991sw} \footnote{The reader should  not confuse these $S$ and $T$ which are both vacuum  polarization functions with the operators $S$ and $T$ in (\ref{eff_lag full2})}:

\be\label{peskin-s}
S = 16 \pi \frac{d}{d q^2} \left[\Pi_{33}(q^2) - \Pi_{3Q}(q^2) \right]\vert_{q^2=0},
\ee 
\be\label{peskin-t}
T = \frac{4 \pi}{s^2 M_W^2}  \left[\Pi_{11}(0) - \Pi_{33}(0) \right],
\ee 
\be
U = 16 \pi \frac{d}{d q^2} \left[\Pi_{11}(q^2) - \Pi_{33}(q^2) \right]\vert_{q^2=0},
\ee 
where $M_W$ is the mass of the $W$, and $s^2\equiv \sin^2{\theta_W}$. The subscripts 1 and 3 refer to the weak $SU(2)$ currents, while $Q$ denotes the electromagnetic current. In practise, only the $S$ and $T$ parameters are relevant for our work because the $U$ parameter is suppressed by an extra factor of the heavy fermion masses. These parameters are  a measure of the size of electroweak breaking, which is parametrized by the breaking scale $y v_H$. On the other hand, they must be suppressed by the mass scale $M$ in the limit $M\to \infty$. Therefore even if one introduces heavy fermions with extremely large couplings, their contributions to the $S$, $T$ and $U$ parameters can become small at least in the limit $M\gg yv_H$. This leaves ample available parameter space within which  extra vector-like fermions are in  agreement with all experimental constraints.

In this section, we analyze how current limits on $S$ and $T$  from precision electroweak fits can be used to obtain constraints on the mass splittings of heavy fermions. To calculate $S$ and $T$, we use exact one-loop expressions for the gauge-bosons self-energies which are valid for all values of the new vector-like fermion masses. 

Our scenario contains three vector-like fermion fields,  one doublet and two singlets of $SU(2)_L$, transforming under the fundamental representation of $SU(n_v)$,
\be
\psi_q =\left(\begin{array}{c}
Q^U \\ Q^D
 \end{array} \right)\;\;\;\;\;\;\; \psi_u= U \;\;\;\;\; \psi_d= D
\ee
with hypercharge $Y$, $Y+1/2$ and  $Y-1/2$, respectively. We will make our computations using the above quark-like fermions. With a simple modification, our results can be readily applied to the case in which the fermions have lepton quantum numbers.  

The full fermionic Lagrangian is given by
\begin{multline}
\lag =    \bar \psi_q( D_\mu \gamma^\mu +m_q) \psi_q +  \bar \psi_u (D_\mu \gamma^{\mu} + m_u)\psi_u  + \bar \psi_d (D_\mu \gamma^{\mu} + m_d)\psi_d +\\+ (y_u \bar \psi_q H \psi_u + y_d  \bar \psi_q H \psi_d +h.c.) 
\end{multline}
where $m_q$, $m_u$ and  $m_d$ are Dirac masses. For simplicity we have assumed that the mass matrix is symmetric and real, but complex masses may be present as well. Finally, the covariant derivative is
\be
D_\mu = \partial_\mu -ig  T_a W_\mu^a - i  g' Y	 B_\mu.
\ee

When the Higgs acquires an expectation value $\langle H\rangle=\left(\begin{array}{c}
0 \\ v
\end{array}\right)$,  off-diagonal mass terms are induced for $\psi_q$, $\psi_u$ and $\psi_d$ 
\be
M_u = \left(\begin{array}{cc}
m_q & y_u \,v\\
y_u\, v  & m_u
\end{array}\right)\;\;\;\;\;\;\;M_d = \left(\begin{array}{cc}
m_q & y_d \,v\\
y_d \,v  & m_d
\end{array}\right)
\ee
The mass matrix $M$ is diagonalized by
\be
\left(\begin{array}{c}
\psi_1 \\ \psi_2
\end{array}\right) = \left(\begin{array}{cc}
c_1 & s_1\\
-s_1 & c_1
\end{array}\right)
\left(\begin{array}{c}
Q^u\\ \psi_u
\end{array}\right)\;\;\;\;\left(\begin{array}{c}
\psi_3 \\ \psi_4
\end{array}\right) = \left(\begin{array}{cc}
c_2 & s_2\\
-s_2 & c_2
\end{array}\right)
\left(\begin{array}{c}
Q^d\\ \psi_d
\end{array}\right)
\ee
where $c_1=\cos \phi_1$, $s_1=\sin \phi_1$, $c_2=\cos \phi_2$, $s_2=\sin \phi_2$ with
\be
\tan 2 \phi_1 = \frac{2 y_u v }{ m_q - m_u}\;\;\;\;\tan 2 \phi_2 = \frac{2 y_d v }{ m_q - m_d}.
\ee
The corresponding eigenvalues are given by
\be
m_{1,2} = \frac12 \left( m_q+m_u \pm \sqrt{(m_q-m_u)^2 +4 y_u^2} \right)
\ee
\be
m_{3,4} = \frac12 \left( m_q+m_d \pm \sqrt{(m_q-m_d)^2 +4 y_d^2} \right)
\ee
 with  the following inverse relations,
\begin{eqnarray}
m_q &=&c_1^2 \,m_1 +s_1^2 \,m_2 = c_2^2 \,m_3 +s_2^2 \,m_4\\
m_u&=& s_1^2\, m_1 +c_1^2 \,m_2 \nonumber\\
m_d&=& s_2^2\, m_3 +c_2^2 \,m_4 \nonumber\\
2\, y_u \,v& =&(m_1-m_2) \,\sin 2\phi_1\nonumber \\
2\, y_d \,v& =&(m_3-m_4) \,\sin 2\phi_2 .\nonumber
\end{eqnarray}

 These expressions now permit the evaluation of the S and T parameters as follows,
\begin{multline}
S= \frac{n_v\,N_c}{3 \pi} \left\{-\frac{c_1^2}{2}(Y+\frac{s_1^2}{2}) \log m_1^2 -\frac{s_1^2}{2}(Y+\frac{c_1^2}{2}) \log m_2^2+\frac{c_2^2}{2}(Y-\frac{s_2^2}{2})\log m_3^2+ \right.\\ \left.+\frac{s_2^2}{2}(Y-\frac{c_2^2}{2})\log m_4^2 +\frac{3 c_1^2 s_1^2}{8}\Pi^{'}(0,m_1,m_2)+\frac{3 c_2^2 s_2^2}{8}\Pi^{'}(0,m_3,m_4)\right\}
\end{multline}
\begin{multline}
T=\frac{n_v\, N_c}{8 \pi s^2 M_W^2}\left\{ c_1^2 c_2^2 \Pi(0,m_1,m_3)+c_1^2 s_2^2 \Pi(0,m_1,m_4)+s_1^2 s_2^2 \Pi(0,m_2,m_4)+\right.\\ \left.
+s_1^2 c_2^2 \Pi(0,m_2,m_3)- c_1^2 s_1^2 \Pi(0,m_1,m_2)-c_2^2 s_2^2 \Pi(0,m_3,m_4) \right\}
\end{multline}
where the functions $\Pi(0,m_1,m_2)$ and $\Pi'(0,m_1,m_2)$ are given by
\begin{multline}
\Pi\left(0,m_1,m_2\right)=\frac{1}{m_1^2 - m_2^2}\left(-m_1^4 + 4 m_1^3 m_2 + 4 m_1^3 (m_1 - 2 m_2) \log{m_1} - 
  4 m_1 m_2^3 + \right.\\ \left.+ 4 m_2^3 (2 m_1 - m_2) \log{m_2} + m_2^4\right)
\end{multline}
\begin{multline}
\Pi^{'}\left(0,m_1,m_2\right)=\frac{2}{9 (m_1^2 - 
   m_2^2)^3}  \left(-12 m_1^3 (m_1^3 - 3 m_1 m_2^2 + 3 m_2^3) \log{m_1} + \right.\\ \left.+
   12 m_2^3 (3 m_1^3 - 3 m_1^2 m_2 + m_2^3) \log{m_2} +\right.\\ \left.+ (m_1^2 - m_2^2)  (2 m_1^4 + 9 m_1^3 m_2 - 16 m_1^2 m_2^2 + 
      9 m_1 m_2^3 + 2 m_2^4)\right).
\end{multline}

In this context, the $S$ and $T$  parameters are a measure of the deviation of the heavy particles from the pure Dirac mass case. 
When the couplings $y_r$ are turned off, the custodial $SU(2)_c$ and isospin $SU(2)_L$ symmetries are restored, and both $S$ and $T$ vanish.
To see this in more detail, we can expand the S and T parameters  in powers of $(m_1-m_2)/(m_1+m_2)$, $(m_3-m_4)/(m_3+m_4)$.
If $m_q=m_u=m_d=M\gg y_u v, y_d v$  we obtain,
\be\label{s-bound}
S= \frac{N_c\,n_v\,v^2\,\left[(11+20 Y)y_u^2+(11-20 Y)y_d^2\right]}{30 \pi M^2}
\ee
\be\label{t-bound}
T = \frac{N_c\,n_v \,v^4\,( y_u^2-y_d^2)^2}{40 \pi\, s_W^2\, M_W^2 \,M^2}.
\ee
where $s_W=\sin\theta_W$ and $M_W$ is the mass of the $W$ boson. Then we see that $S,T\to 0$ when $y_u,y_d\to 0$, i.e.,  $m_2\to m_1$ and $m_4\to m_3$ . The corresponding formulas for the lepton-like fermion case can be obtained by substituting $y_u\to 0$, $y_d\to y_l$, $N_c\to 1$ in equations (\ref{s-bound}) and (\ref{t-bound}).
\begin{figure}[th!]
\begin{center}
\includegraphics[width=0.4\textwidth]{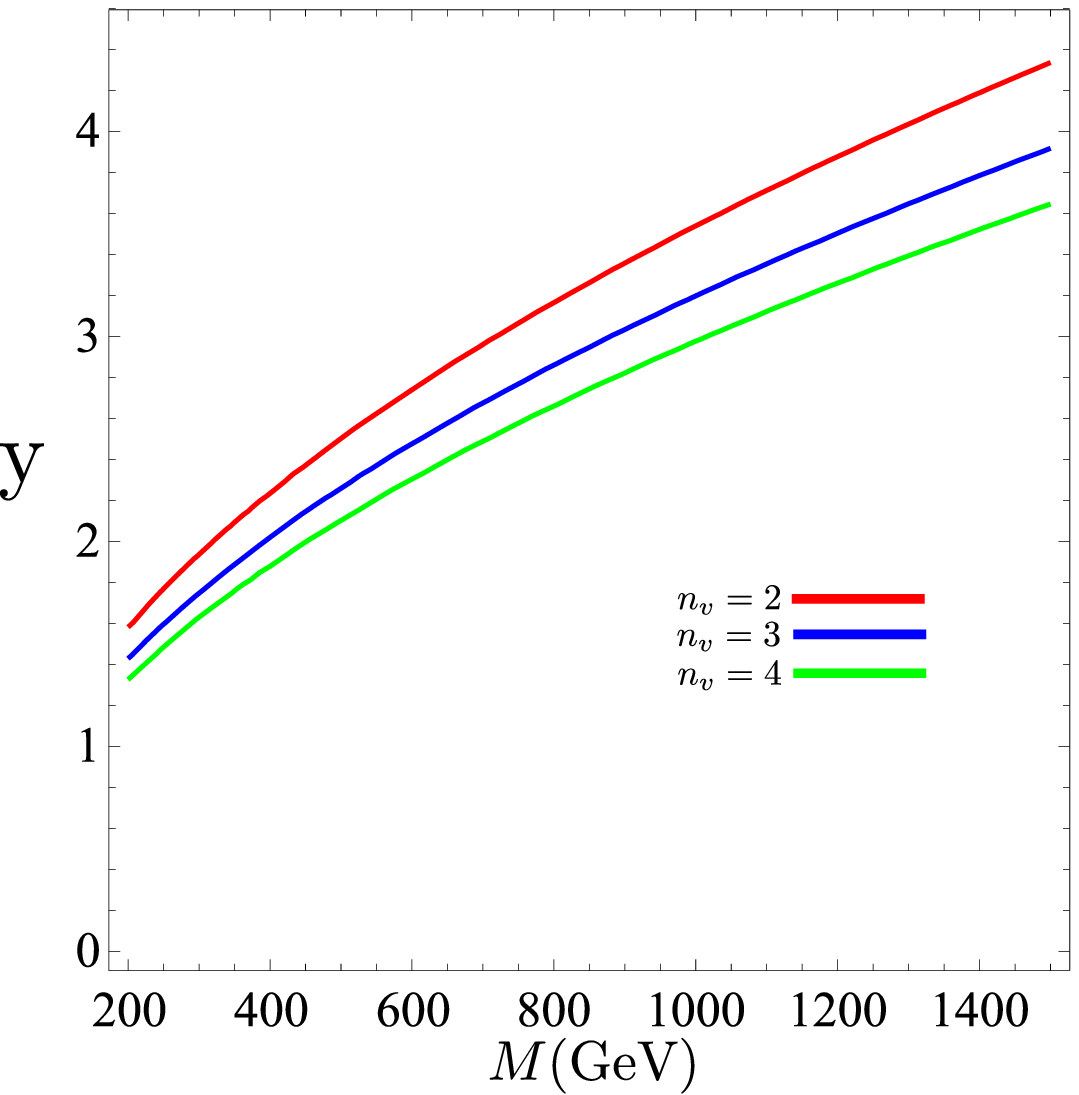}
\includegraphics[width=0.4\textwidth]{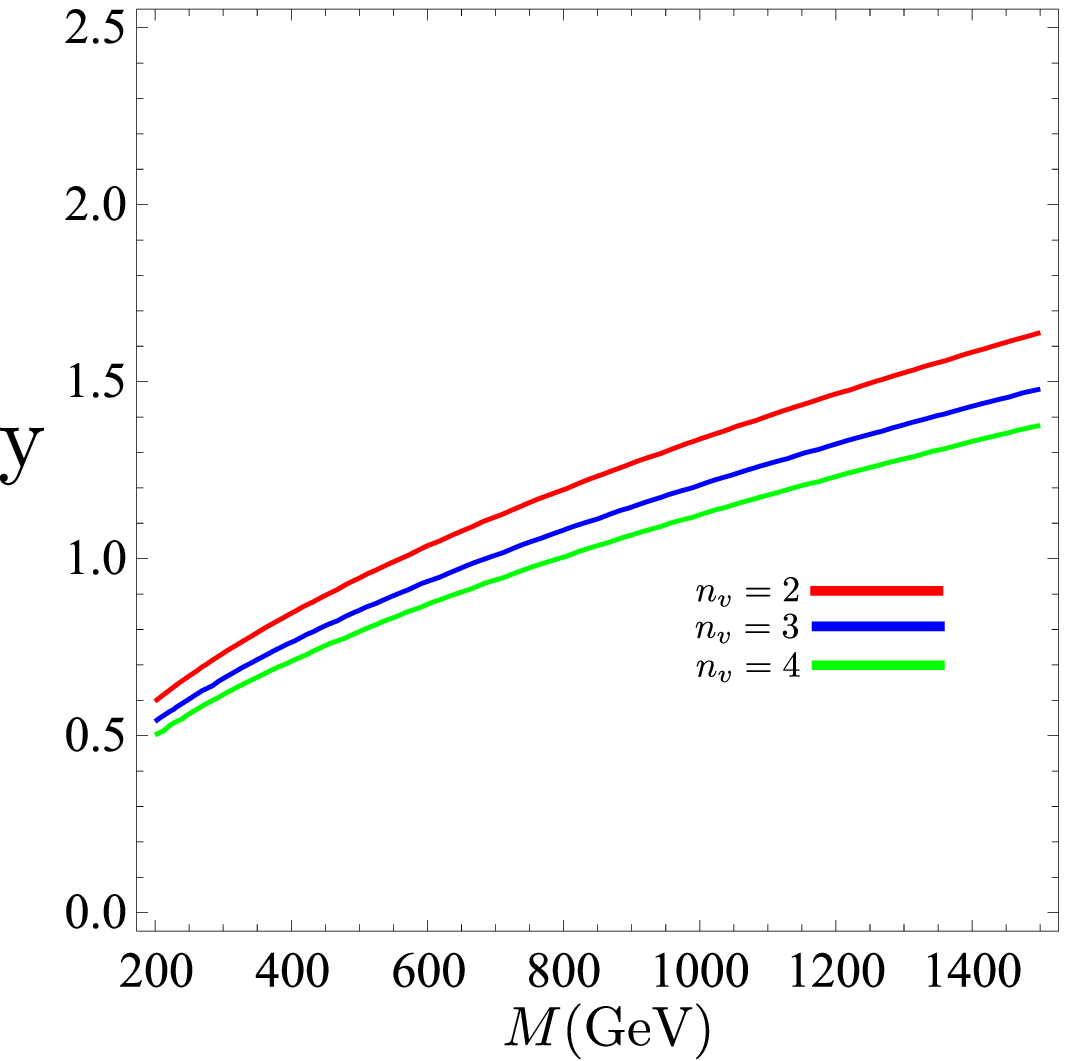}\end{center}
\caption{The bounds at $95\%$ CL on the $(M,y)$ parameters from constraints on the oblique parameters $(S,T)$ for $n_v=2,3,4$
and two different regimes: $\rho_r\approx 1 \,(y=\sqrt{7} y_l)$ (Left panel), and  $\rho_{\bar d}= \rho_{\bar u}=\rho_{q}\gg \rho_l=\rho_{e} \approx 1 (y=y_l)$ (Right panel).  The upper-left region is excluded in these plots. }\label{plot-bounds}
\end{figure} 

 Fits of the combined electroweak data provide constraints on the $S$ and $T$ parameters and have been obtained in many places. Here we use the results from the PDG fits. The standard model is defined by $(S,T) =(0,0)$ with $m_t=170.9\GeV$ and $m_H=115\GeV$. The best fit to data  is (without fixing $U=0$)
\be
S = -0.10\pm 0.10
\ee
\be
T= -0.08\pm 0.11.
\ee
These equations then imply the bounds $S\leq 0.10$ and $T\leq 0.13$ at  $95\%$ CL.
From the experimental data, we can readily find the constraints on the parameters of the model. For example, for lepton-like fermions, if we take $Y=-1/2$ and $m_1=1\TeV$, then $S\sim 0.01 y_l^2 n_v $ and $T\sim 0.08 y_l^4 n_v$. Notice that the small value of  $S$   leads to only mild constraints on $y_l$ and $n_v$. By constrast, the parameter $T$ provides the most stringent bounds. 

There is, however, one important effect that can mitigate the contribution to $T$. From (\ref{t-bound}), we see that electroweak corrections to the $T$ parameter are insensitive along the $y_u=y_d$ direction in the parameter space, where the isospin symmetry is restored. In this case, the only constraints on the $y_u$ and $y_d$ contributions to $y$ are given by the $S$ parameter, which, as we have just seen, are less stringent.

In figure \ref{plot-bounds} we show the bounds on $y$  at the $95 \%$ C.L. from the $X$ particles in table \ref{tab reps} as a function of the mass scale $M$ and for $n_v=2,3,4$. For illustration, we have considered   $y_u =  y_d = y_l$ and two different regimes: the degenerate case, $\rho_r\approx 1$, (Left panel) and the non-degenerate case, $\rho_{\bar d}= \rho_{\bar u}=\rho_{q}\gg \rho_l=\rho_{e}\approx 1$, (Right panel). In this case, one has $y=\sqrt{7} y_l$ and $y=y_l$ for the degenerate and non-degenerate cases, respectively, (cf. (\ref{y})). Since $y_u$ and $y_d$ are unconstrained by $T$ along the $y_u=y_d$ line, and the limits from $S$ are less severe, the only bound we can obtain on $y$ comes  from the most stringent bound on $y_l$ due to the $T$ parameter. 
We then see that typical values $M\simeq 1, 1.5\TeV$ and $y\simeq 1$ are permitted for $n_v=2$. For larger values of $n_v$, the allowed split between the mass eigenstates is smaller, but still large enough that a coupling $y\simeq 0.5$ is not unreasonable.

To summarize, we  conclude that for a sufficiently large range of the parameters the fit to electroweak observables is in agreement with the existence of  extra vector-like fermions. This parameter space is characterized by
\be\nonumber \label{y_limits}
\rho_r\approx 1:\,\,\,y \lesssim 1.2-1.6
\ee  
\be
\rho_{\bar d}= \rho_{\bar u}=\rho_{q}\gg \rho_l=\rho_{e}:\,\,\,y\lesssim 0.6-1.2
\ee
together with the current direct search limits
\be\nonumber
M_l,M_{\bar e}\gtrsim 200\GeV
\ee
\be
M_q,M_{\bar u},M_{\bar d}\gtrsim 250\GeV.
\ee

\section{Numerical analysis}
\setcounter{equation}{0}
An interesting feature of the pure-glue hidden valley is that the v-glueballs can have many decay modes, depending on their quantum numbers and on the values of the various parameters. To give a  general survey of decays is beyond the scope of this work, since a more detailed treatment would have to incorporate precise values of the matrix elements which at present are unknown. However, a few examples  are useful to illustrate the main qualitative features of v-glueball decays.

\subsection{Decay patterns of v-glueballs}

 The relevant parameter space  consists of the mass scale $M$, the $0^{++}$ mass $m_0$, the coupling $y$ and the Higgs mass  $m_H$. Here we present our results in terms of $m_0$ which is more transparent than the  confining scale $\Lambda_v$, since $m_0$ is the relevant parameter for LHC studies. To simplify the discussion in the SM, it is convenient to assume that the Higgs boson is SM-like and fix its mass to be $m_H=120$ (``low mass'' range) or $m_H=200\GeV$ (``high mass'' range). Moreover, since the branching ratios depend on $M$ and $y$ only through the combination $yM$,  our problem is reduced  to a two-dimensional parameter space described by $yM$ and $m_0$.
 
 For our estimates below,  we will use the lattice results~\cite{Morningstar2} 
 \footnote{ Here the coupling constant $4\pi \alpha_v$ is included alongside   ${\bf F_{0^{++}}^S}$ and ${\bf F_{0^{-+}}^P}$  to make them renormalization invariant so that there is no question at which point $4\pi \alpha_v$ is normalized. On the contrary, ${\bf F_{2^{++}}^T}$ is renormalization invariant as is, since it is the matrix element of the energy-momentum tensor which is known to be scale invariant. Also, since the values reported  in \cite{Morningstar2} are not expressed in a continuum renormalization scheme, we have converted $ g^2 {\bf F_{2^{++}}^T}$ to ${\bf F_{2^{++}}^T}$ using the value of the lattice parameter $\beta=6/g^2=3.2$ as an approximation to the continuum limit.
}
 \be\label{decay_constants}
  4\pi \alpha_v {\bf F_{0^{++}}^S} = 3.06 m_0^3, \;\;\;\;  {\bf F_{2^{++}}^T} = 0.03 m_0^3\, (n_v/3) ,\;\;\;\; 4\pi \alpha_v {\bf F_{0^{-+}}^P} = 0.83 m_0^3 .\ee
We also need estimates of the other v-glueball decay constants and transition matrix elements, which at this point are unknown. A reasonable  educated guess is that they are of order $ {\bf F_{2^{++}}^T}$, which is the only known decay constant that is largely independent of the size of the v-glueballs. We can also guess ${{\bf M_{1^{--}0^{++}}^\Omega}}\sim 1/{\bf F^{S}_{0^{++}}}$, as it would be true for pion emission. In the following, we will therefore assume 
\be \label{approx-me}
{\bf F_{2^{\pm+}}^L}= \frac{n_v}{3}{\bf M^{S(i)}_{2^{++} 0^{++}}} m_0 \simeq{\bf F_{2^{++}}^{T}}, \;\;\;\; \;\;{\bf M^{S(i)}_{1^{--} 1^{+-}}} m_{1^{+-}}= \sqrt{\frac{n_v}{3}}{{\bf M_{1^{--}0^{++}}^\Omega}}\simeq \frac{n_v}{3} m_0^6/{\bf F^{S}_{0^{++}}}.
\ee
Later in this section, we will comment on how large deviations from this guess might affect our estimates.

\begin{figure}[h!]
\begin{center}
\includegraphics[width=0.32\textwidth]{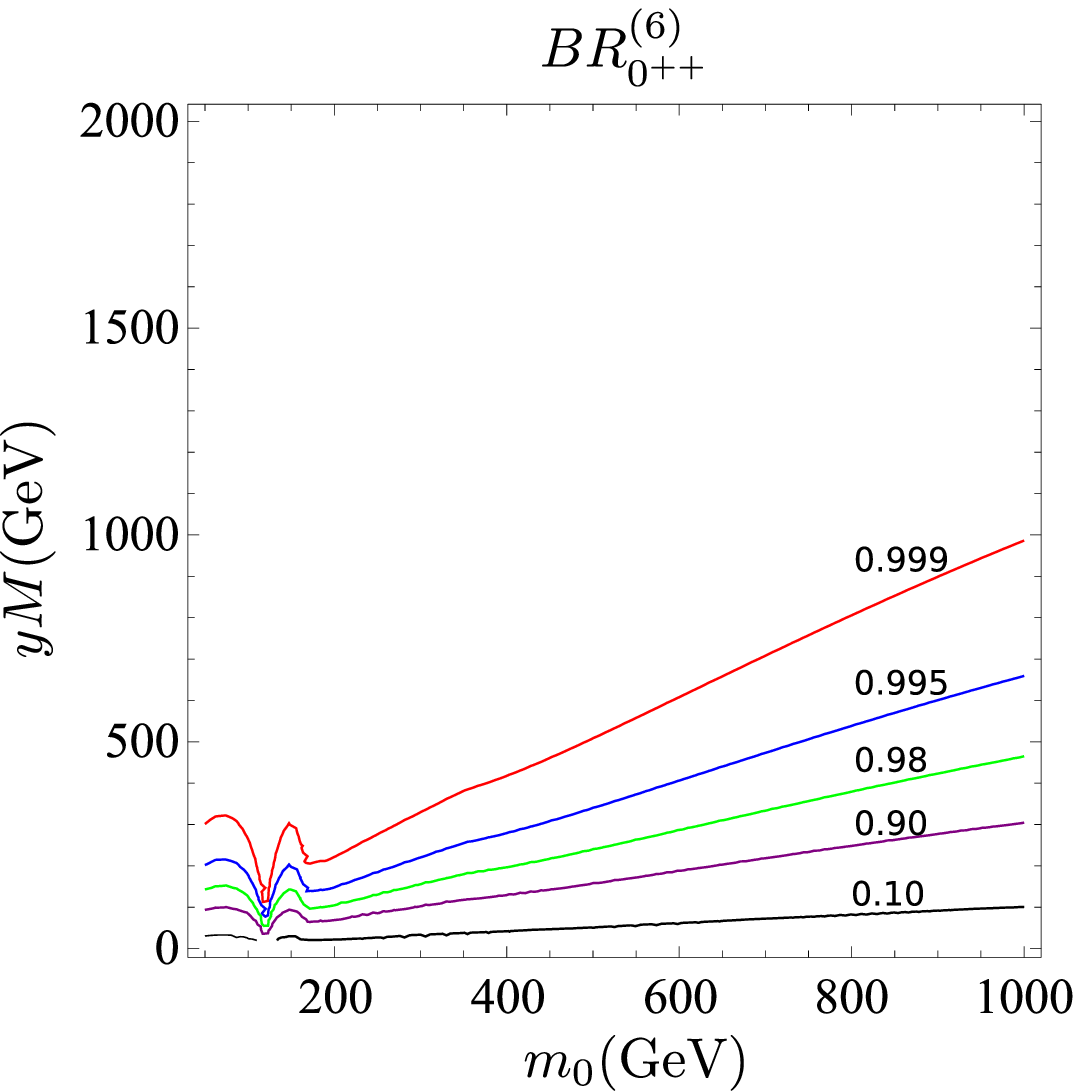}
\includegraphics[width=0.32\textwidth]{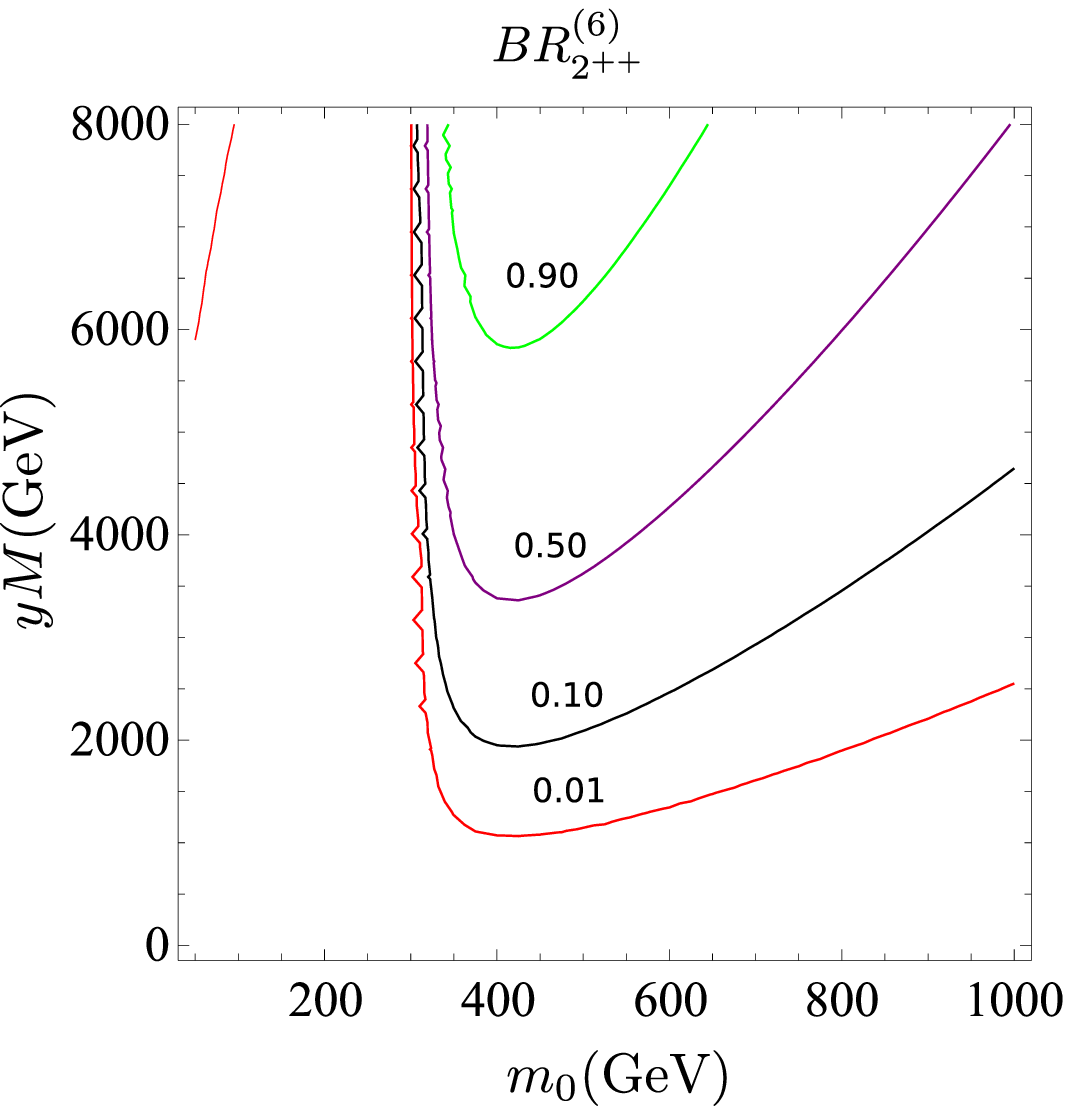}
\includegraphics[width=0.32\textwidth]{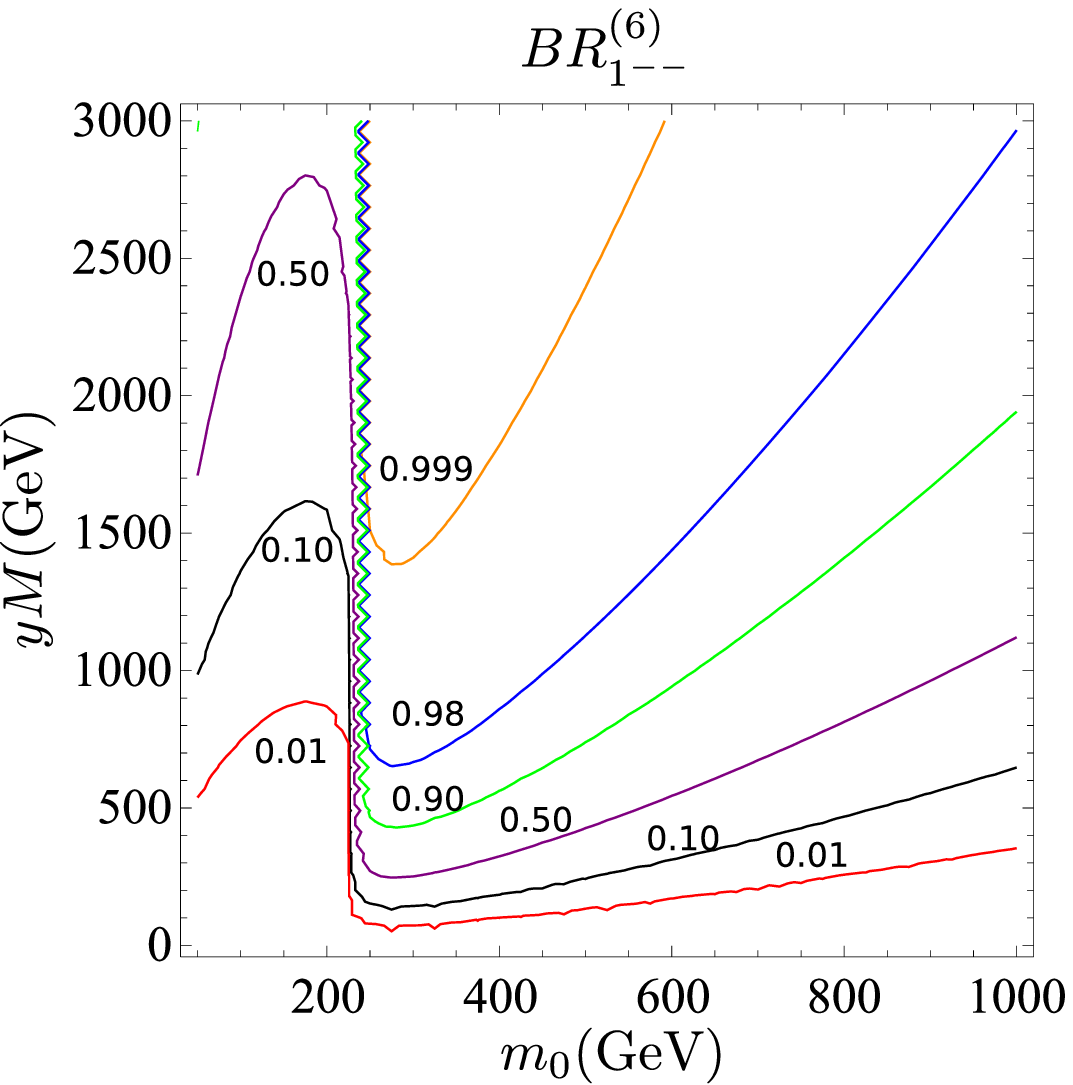}
\end{center}
\caption{Curves of constant branching ratio $BR^{(6)}$ in the parameter space $(m_0, yM)$  for various representative states and $m_H=120\GeV$. Left: $0^{++}$, Center: $2^{++}$, Right: $1^{--}$. }\label{bounds1}
\end{figure}

\begin{figure}[h!]
\begin{center}
\includegraphics[width=0.32\textwidth]{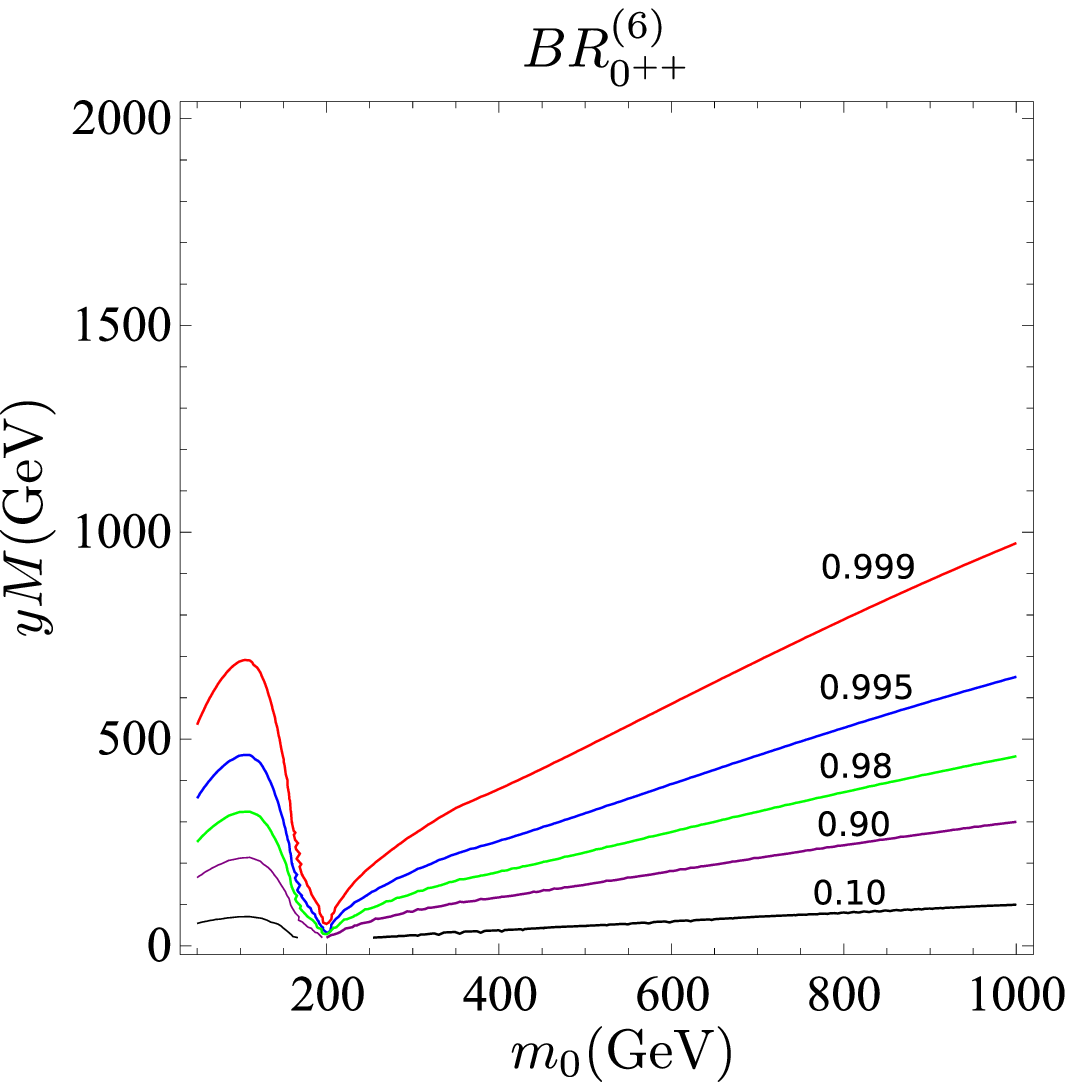}
\includegraphics[width=0.32\textwidth]{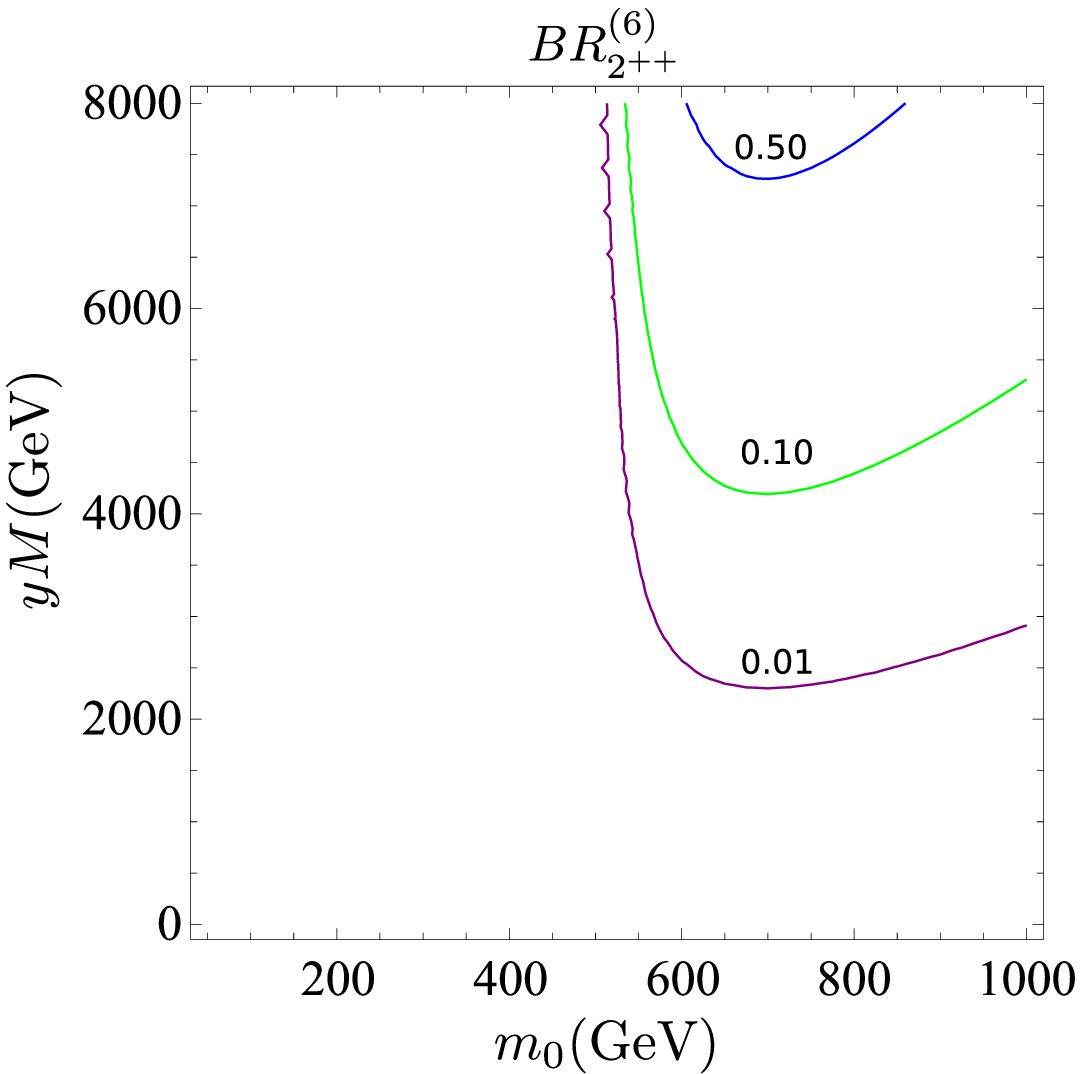}
\includegraphics[width=0.32\textwidth]{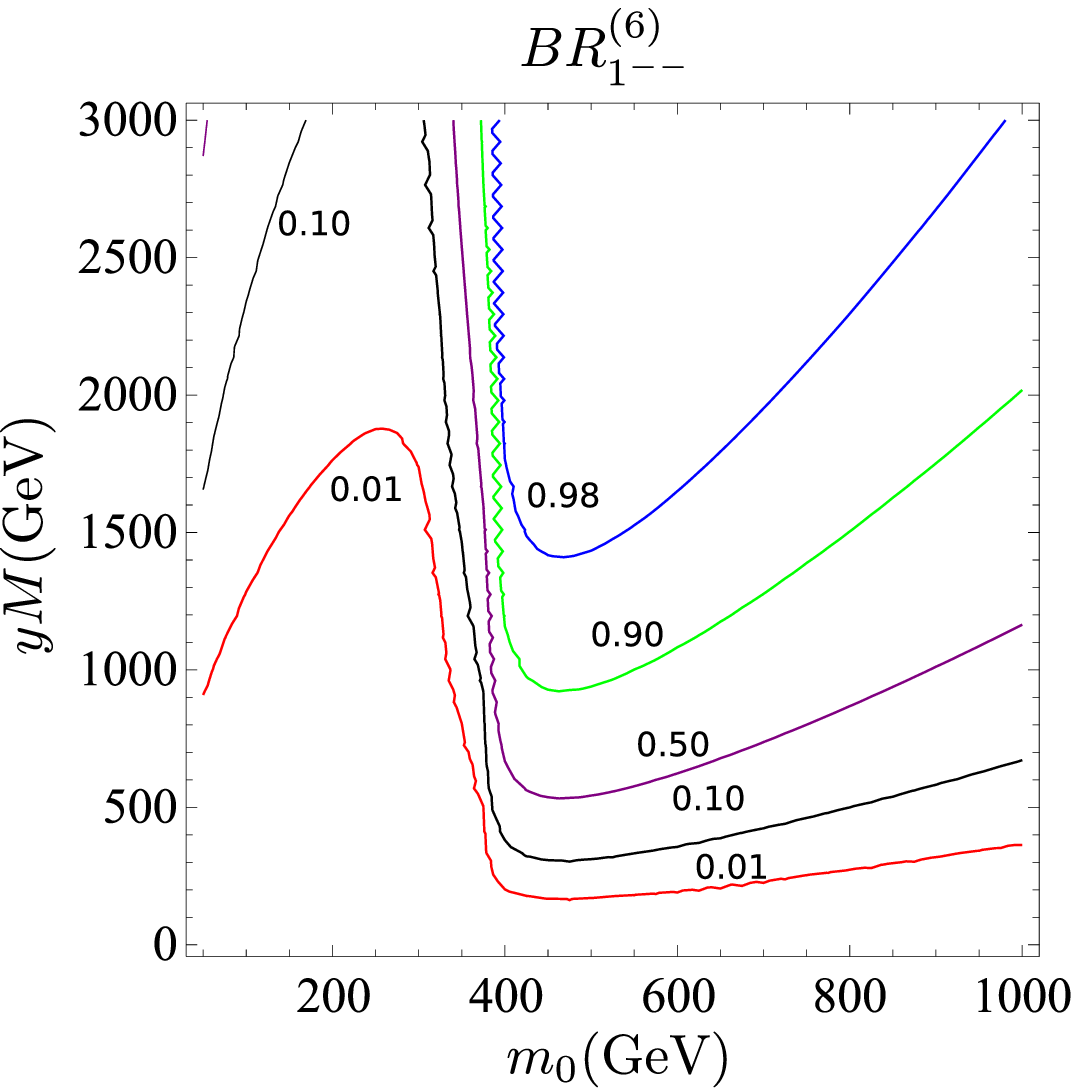}
\end{center}
\caption{Same as figure \ref{bounds1} for $m_H=200\GeV$. }\label{bounds2}
\end{figure}

Using the decay rates expressed in section 3, we identify $\Gamma^{(6)}$ and $\Gamma^{(8)}$ as the summed contributions
to the decay rates from dimension-six and dimension-eight operators, respectively.
The corresponding branching ratios are denoted by $BR^{(6)}$ and $BR^{(8)}$, satisfying $BR^{(6)}+BR^{(8)}=1$.
To illustrate the dependence of the branching fractions on $yM$ and $m_0$, we present in figures \ref{bounds1} and \ref{bounds2} contours of constant $BR^{(6)}$  in the $m_0$ vs $yM$ plane for various choices of v-glueball states and $m_H=120\GeV$, $200\GeV$.
In figure \ref{bounds1}, we see that the values of $BR^{(6)}_{0^{++}}$ are large over most of the parameter space plane, except  at very low $yM$. By contrast,  the values of  $BR^{(6)}_{2^{++}}$  are small, except for a small region at large $yM$.
 The most interesting decay pattern is found in the $1^{--}$ v-glueball. We see that $BR^{(8)}_{1^{--}}$ typically dominates when $m_{1^{--}}-m_{1^{+-}}<m_H$ and the emitted Higgs boson is off-shell, whereas $BR^{(6)}_{1^{--}}$ dominates  when $m_{1^{--}}-m_{1^{+-}}>m_H$ and the emitted Higgs boson is on-shell. An analogous behaviour is found for $m_H=200\GeV$, as demonstrated in figure \ref{bounds2}.

The values of $yM$ determine to a large extent the decay pattern of v-glueballs with three distinctive energy regions: $(i)\; yM \ll 1\TeV$, $(ii) \; yM \gg 1\TeV $ and $(iii) \; yM \approx 1\TeV$. Recall that these ranges of parameters are further constrained by existing experimental limits, as described in  section 4.  As will be discussed below, the common feature of all these different regimes is the great diversity of decay modes and lifetimes, even for a given choice of parameters.

\vspace{3mm}

$(i) \;yM \ll 1\TeV$. For sufficiently small $yM$, the contribution of the dimension-six operators can be effectively disregarded and, hence, the dimension-eight operators have
the largest effects on v-glueball decays.
As shown in \cite{Juknevich:2009ji}, annihilation decays dominate the $0^{++}$, $2^{++}$, $0^{-+}$ and $2^{-+}$ v-glueballs. Their branching ratios are dominated by decays to $gg$, with decays to $\gamma\gamma$ having a branching fraction of $\sim 0.4\%$, assuming the $X$ fields form complete $SU(5)$ multiplets of equal mass.  If the colored particles are much heavier than the uncolored particles, then decays to electroweak bosons and photons can dominate. Most other states decay by radiatively emitting a photon, or a $Z$ boson. In addition, the $1^{--}$ state can also decay to standard model fermions via an off-shell $\gamma$ or $Z$. The $3^{++}$ state is special in that its three-body decay mode $3^{++}\to 0^{++}gg$ could be of the same rate as the radiative $3^{++}\to 1^{+-}\gamma$ decay. The clearest signatures for this regime are likely to be the two-photon resonances from the prompt annihilation decays of $C$-even v-glueballs. In addition, for part of the parameter space, the v-glueballs are rather long-lived particles, so that they produced displaced vertices in the detectors, which can be an experimentally challenging, but certainly important signature.

\vspace{3mm}

$(ii) \; yM\gg 1\TeV   $. In this case, the Higgs couplings to v-glueballs dominate over all other couplings. As a result, the decay pattern is relatively simple.
In the $PC=++$ sector,  we find that the $0^{++}$ v-glueball annihilates directly into pairs of standard model particles via  $0^{++}\to h^*$ , with the same final states and branching fractions of a Higgs boson with mass $m_0$. For $m_0>2m_h$, the mode $0^{++}\to hh$  opens up, in addition to $0^{++}\to h^*$, contributing $20-25\%$ to the total width with the decays into weak boson and top quark pairs accounting for the remaining $75-80\%$. The $2^{++}$ v-glueball  decays predominantly to the lighter $0^{++}$ by radiatively emitting a Higgs boson. In the $PC=-+$ sector the lightest states are the $0^{-+}$ and $2^{-+}$ v-glueballs. The $2^{-+}$ v-glueball  decays in a similar way to the $2^{++}$, with the rate dominated by $2^{-+}\to 0^{-+} h$. Although the $2^{-+}$ v-glueball is heavier than the $0^{++}$, the  $2^{-+}\to 0^{++} h$ decay is not allowed, due to parity and conserved angular momentum. The  $0^{-+}$ v-glueball is a special case. Without explicit breaking of parity,  the $0^{-+}$ v-glueball  decays  slowly into $gg$ via dimension-eight operators.

 Turning to the $PC=+-$ sector, we find that most of the states can decay  by emission of a Higgs boson, through processes of the form $\Theta_\kappa\to \Theta_{\kappa'} h$, where $\Theta_\kappa$, $\Theta_{\kappa'}$ denote two v-glueballs with given quantum numbers. Since C-changing transitions are not allowed by the $S$ operator, the v-glueballs in that sector typically  undergo a cascade decay, radiating a Higgs boson at each step, which ends at the lightest $1^{+-}$ state. The v-glueballs in the $PC=--$ sector decay in a similar manner, with the lightest $1^{--}$ v-glueball decaying to the $1^{+-}$ v-glueball with the emission of a Higgs boson. Since  the $1^{+-}$ v-glueball is not permitted to decay via dimension-six operators,  its dominant decay modes are photon-radiative transitions to  $C$-even v-glueballs,  as in $(i)$. 

We should note that this theoretical regime may be difficult to access in practice. For the $2^{++}$ v-glueball (and hence the $2^{-+}$), this requires  $yM\gtrsim 10\TeV$ (see figures \ref{bounds1} and \ref{bounds2}). As shown in (\ref{y_limits}), the heavy mediators can at most have Yukawa couplings $y\sim 1$ in order to avoid potentially dangerous electroweak corrections.  With  $M\sim {\cal O}(10\TeV)$, the rate for production of v-glueballs ($\propto \sigma_{gg\to XX}\ll 1 \rm fb$) is rather small, rendering this region of parameter space experimentally inaccessible to the future experiments at the LHC. 

\vspace{3mm}

$(iii) \; yM \approx 1\TeV$.  For Higgs couplings with intermediate strength, the v-glueball decay pattern is more complicated because the couplings to the gauge bosons become important, leading to a interesting interplay between dimension-six and dimension-eight operators.  From the point of view of prospective experiments at the LHC, this is also the most interesting regime, due to the diversity of v-glueball decay channels and variability of lifetimes as well.

\begin{figure}[tb]
\begin{center}
\includegraphics[width=0.44\textwidth]{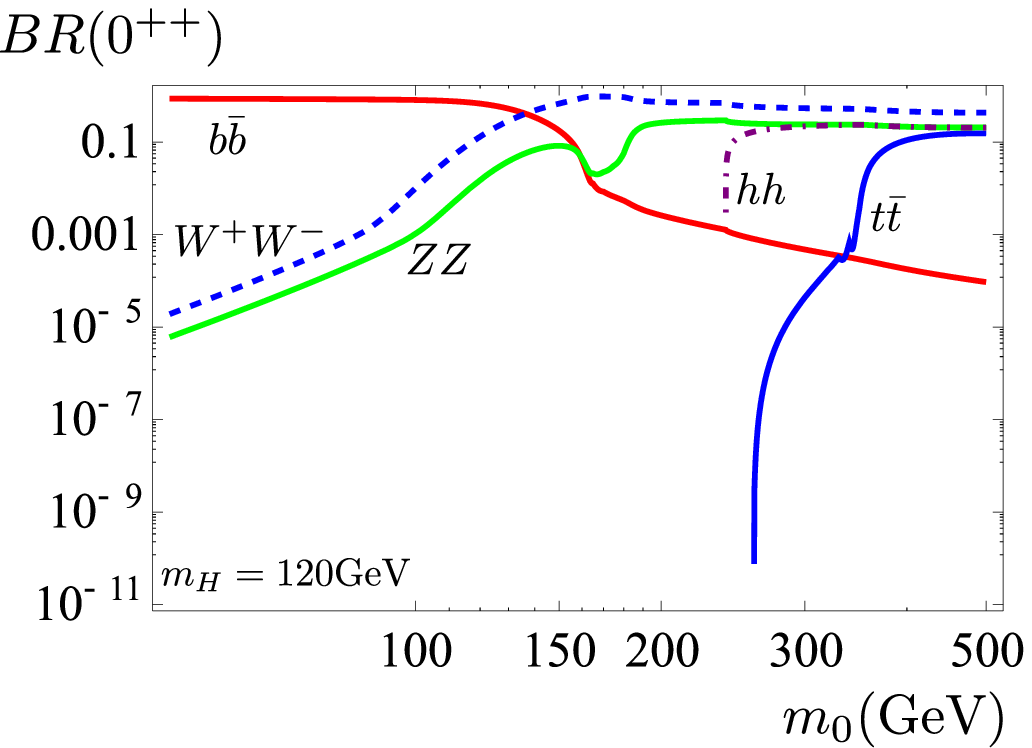}
\includegraphics[width=0.44\textwidth]{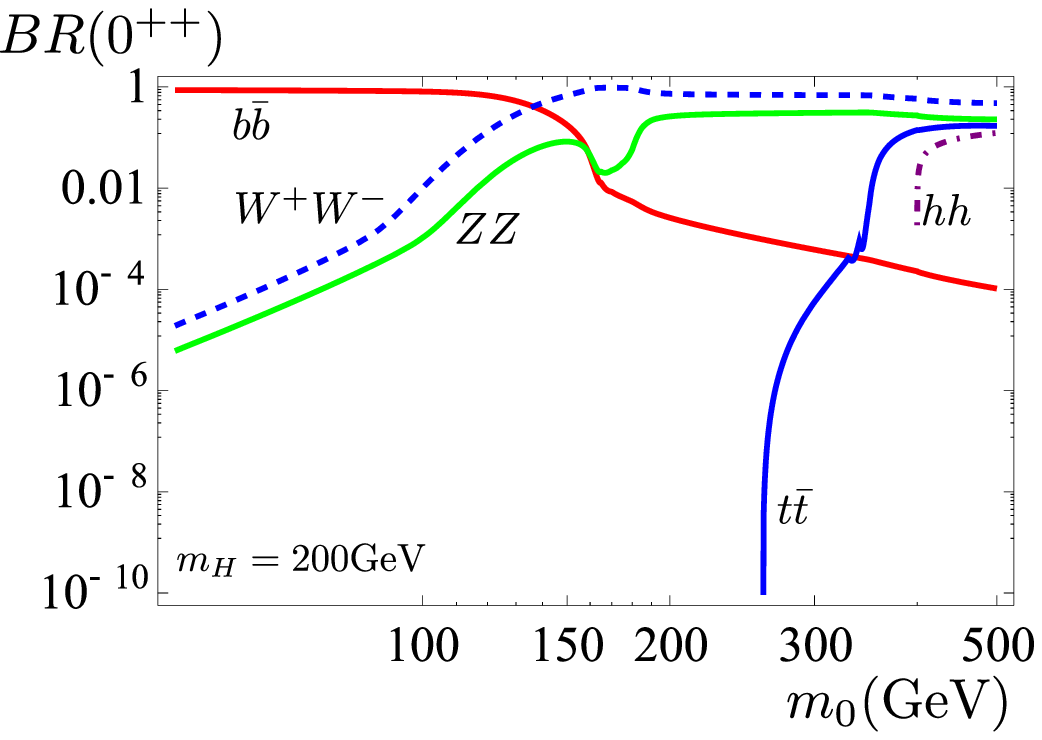}
\end{center}
\caption{The branching ratios of the $0^{++}$ v-glueball as a function of $m_0$ for $ yM \sim 1\TeV$. Left Panel: $m_H=120\GeV$. Right Panel: $m_H=200\GeV$. For clarity, only the main decay modes are shown. } \label{plot-br0}
\end{figure}

The $0^{++}$ v-glueball still decays predominantly to pairs of standard model particles via $0^{++}\to h^*$ or, if kinematically allowed, $0^{++}\to hh$. The branching ratios for the main decay channels of the $0^{++}$ v-glueball are shown in figure \ref{plot-br0} for $yM=1\TeV$ in the cases of $m_H=120\GeV$ and $m_H=200\GeV$.  

For the $2^{++}$ and $2^{-+}$ v-glueballs, the dominant decay mode is $gg$. This is because the $2^{\pm+}\to 0^{\pm+}h(h^*)$ decay is phase-space suppressed by a $10^{-8}-10^{-3}$ factor. For $m_{2^{++}}-m_{0^{++}}>m_H$,  the phase space suppression is smaller $\sim 0.01$, but in this regime the $(m_{2^{++}}/M)^4$  suppression of $D=8$ operators is inefficient, rendering the $D=6$ operators subdominant.

The dominant decay modes of the $1^{--}$ v-glueball are  transitions with a Higgs boson in the final state, $1^{--}\to   1^{+-} h$, for $m_{1^{--}}-m_{1^{+-}}>m_H$ or, photon-radiative decays to the lighter v-glueballs in the $C$-odd sector for $m_{1^{--}}-m_{1^{+-}}<m_H$.  The decays of the other v-glueballs in the $C$-odd sector proceed in a similar fashion. In addition, the decay of the $1^{--}$ v-glueball  into fermion pairs can play a significant role for $m_{1^{--}}-m_{1^{+-}}<m_H$.

Since the $1^{+-}$ and $0^{-+}$ v-glueballs cannot decay via their couplings to the Higgs boson,  their dominant decay modes are $gg$ for the $0^{-+}$ v-glueball and radiative transitions to the $C$-even states with emission of a photon for the $1^{+-}$ v-glueball.

The $3^{++}$ state is more complicated. With many contributing decay channels and unknown form factors, it seems impossible to estimate which decay mode is dominant. Indeed, a simple estimate suggests that the $3^{++}\to 0^{++}gg$, $3^{++}\to 1^{+-}\gamma$ and $3^{++}\to 0^{-+}h$ decays are all  of the same order.

\vspace{3mm}

Summarizing all three cases, we expect the v-glueball decays to produce any of the kinematically allowed final states of the Higgs boson, such as $b\bar b $, $\tau^+\tau^-$, $W^+ W^-$, etc, as well as possible photons, both singly and in pairs, and gluon pairs. If kinematically allowed,  multiple production of standard model Higgs bosons from cascade decays can also proceeds with a sizable rate.
A typical final visible state  would then be of the form $b\bar b b\bar b$, $W W W W b\bar b$, $b\bar b \tau^+\tau^-$, $b\bar b gg$, $b\bar b \gamma\gamma$,  and so on.

\subsection{Lifetimes}

\begin{figure}[tb]
\begin{center}
\includegraphics[width=0.32\textwidth]{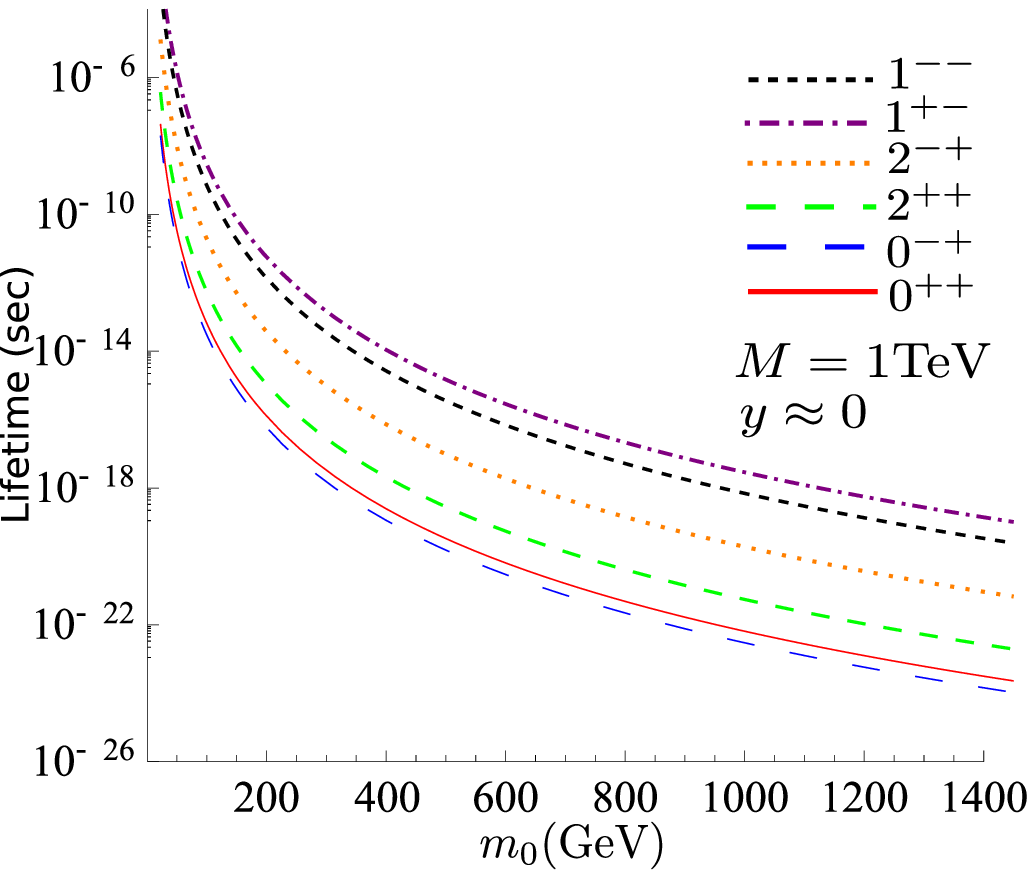}
\includegraphics[width=0.32\textwidth]{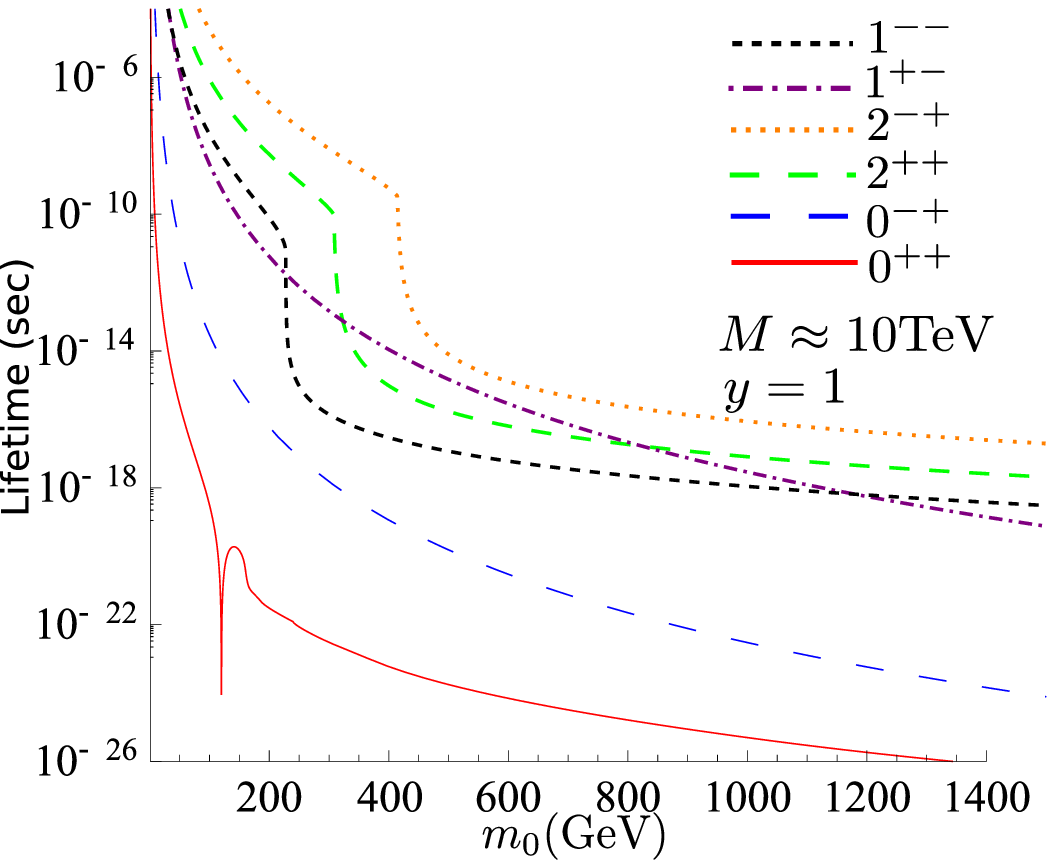}
\includegraphics[width=0.32\textwidth]{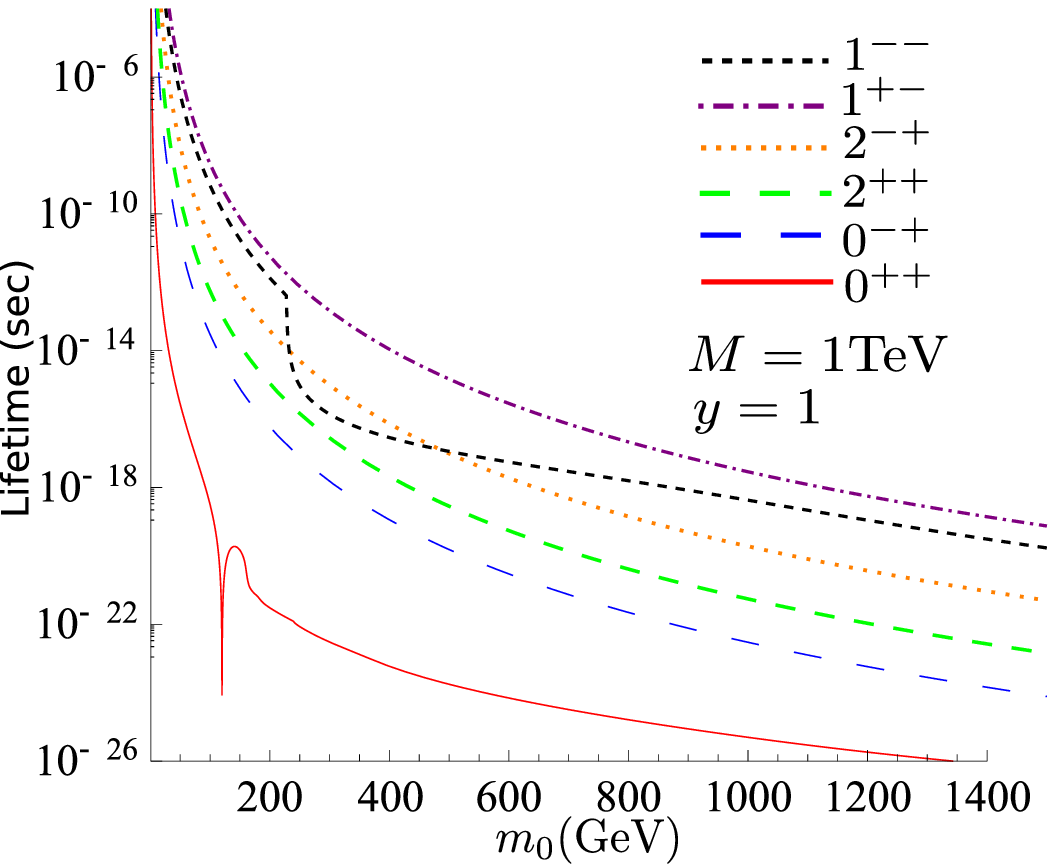}\end{center}
\caption{Lifetimes of the v-glueballs as a function of the v-glueball mass scale $m_0$ for three representative regimes: $yM\approx 0$ (Left panel), $yM\approx 10\TeV$ (Middle panel), and $yM\approx 1 \TeV$ (Right panel). } \label{lifetimes}
\end{figure}

Meanwhile, with so many v-glueball states and decay channels, the lifetimes of the v-glueballs can vary over many orders of magnitude. This is already clear from the fact that the lifetimes are very sensitive to both $M$ and $m_0$. In our approach, the large contributions of the dimension-six operators  also suggest a  wider spread of the lifetimes than that found when the Higgs couplings are absent, as in~\cite{Juknevich:2009ji}.
In figure \ref{lifetimes}, we plot the lifetimes of some of the v-glueballs as a function of $m_0$ for the three representative regimes studied in the last subsection. We see that for $yM\approx 1-10\TeV$ the various v-glueball states have lifetimes that typically span 5-6 orders of magnitude, for any given choice of the parameters. 

An important consequence of the large spread in the lifetimes is that there is a significant probability that one or more of the v-glueballs will often decay a macroscopic distance away from the primary interaction vertex. Only one of these states needs to be both long-lived and frequently produced to provide a strong signature of new physics.

If production rates are substantial, displaced vertices are expected for average decay lengths\footnote{ Note that the actual v-glueball decay length $l$ in the laboratory frame, distributed according to $P(l)\propto e^{-l/\gamma c\tau}$ with the Lorentz boost factor $\gamma=1/\sqrt{1-\beta^2}$ , may be actually much smaller than the average decay length $c\tau$. So even if  $c\tau$  far exceeds the dimensions of the detector, there is still a  significant chance that these v-glueballs will be observable, providing the rates are substantial.} $c\tau$ of the order $10^{-4}-100\, \rm{ m}$,  resulting in lifetimes of $10^{-12}-10^{-6}\, \rm{ sec}$. Outside this range, the displaced-vertex signature is no longer present, either because the lifetime is much longer than $10^{-6}$ sec, so that a typical v-glueball produced at the LHC will escape the detectors, or because the decays are prompt and the v-glueball decay lengths cannot be resolved. 

As shown in figure \ref{lifetimes}, displaced vertices could be observed for $10\GeV \lesssim m_0\lesssim 200\GeV$ and $0\lesssim yM\lesssim 1 \TeV$ or for $10\GeV \lesssim m_0\lesssim 400\GeV$ and $y M\gg 1\TeV$. As long as $m_0$ becomes greater than about $400\GeV$, all v-glueball decays will be prompt and the displaced-vertex signature will be absent.   In the case $yM\approx 0$ the  $1^{+-}$ and $1^{--}$ v-glueball lifetimes reach values high enough for displaced vertices for $m_0$ between $50\GeV$ and $150\GeV$. The $0^{++}$, $0^{-+}$, $2^{++}$ and $2^{-+}$  v-glueballs may also decay with displaced vertices in the very low mass range $m_0\lesssim 70\GeV$. On the other hand, in the case $yM \approx 10\TeV$ the spread in the lifetimes  is two or three orders of magnitude larger, so even larger v-glueball masses $150\GeV <m_0<400\GeV$ may allow for displaced vertices. This is the case of the $2^{++}$ and $2^{-+}$ v-glueballs, as shown in the middle panel of figure \ref{lifetimes}. Their lifetimes are so long that these v-glueballs may escape the detectors for $m_0\lesssim 100\GeV$. In the low mass range $50\GeV <m_0<200\GeV$ other v-glueballs such as the $1^{+-}$ and $1^{--}$ also have a significant chance to decay with displaces vertices. In this case, the $0^{++}$ and $2^{++}$ v-glueballs will remain short-lived over most of the parameter space down to masses of order $50\GeV$ where displaced vertices may start to occur for these states as well. Finally,  the case  $yM\approx 1\TeV$ shown in the right panel of figure \ref{lifetimes} is an example of an intermediate situation, with the $0^{++}$ v-glueball being short-lived, except for very-low masses $m_0\lesssim 50\GeV$, and at least five states having a chance to decay with displaced vertices in some part of the parameter space: the  $1^{+-}$ and $1^{--}$ v-glueballs for $ m_0\lesssim 200\GeV$, and the  $2^{++}$, $2^{-+}$ and $0^{-+}$ v-glueballs for 
$ 20\GeV \lesssim m_0\lesssim 100\GeV$. 


From the analysis above it is quite evident that long-lived resonances are a common feature of pure-glue hidden valleys for an ample range of parameters.
Nonetheless, detecting long-lived  particles presents several experimental challenges. Displaced jets, though they have no standard model background,  suffer from substantial detector backgrounds from secondary vertices inside the detector material. On the other hand, displaced leptons are technically easier, and have much less background, but branching ratios to leptons are usually quite small. Displaced photons also present several experimental challenges, since one has to detect not only the position where the photons enter the electromagnetic calorimeter but also their angle of incidence.  Thus the issue of detecting displaced vertices is not straightforward, and requires novel analysis strategies.

Interestingly enough, the v-glueball masses can be as low as a few GeV's and at the same time the lifetimes could be short enough that a few v-glueball events could be observed at the LHC. Specifically, let us examine closely the case of the scalar $0^{++}$ v-glueball. From (\ref{wid0}) we see that for $X$ particles of order $1\TeV$ the lifetime of the $0^{++}$ v-glueball is of order
\be
 c\tau \sim 1{\rm cm} \left(\frac{M}{1\TeV}\right)^4\left(\frac{20\GeV}{m_0}\right)^7 \left(\frac{5\GeV}{m_b}\right)^2 \left(1-\frac{4 m_f^2}{m_0^2}\right)^{-1.5}.
\ee
We see  that for $m_0\gtrsim 10\GeV$ the $0^{++}$ state can decay inside the detector. It is important to remark that even for $c\tau$ well above $10$ meters, a significant  fraction of the v-glueballs will still decay inside the detector. If detector backgrounds are sufficiently low, and production cross-section is substantial, discovery would also be possible, in principle, for states in the low-mass range $1\GeV\lesssim m_0\lesssim 10\GeV$.

\subsection{Uncolored $X$ particles}

If there is a substantial hierarchy between the colored $X$ particles and the uncolored $X$ particles, the latter may dominate v-glueball decays. Decay rates to a pair of SM gluons in the $C$-even sector are proportional to the coefficients $\chi_3^2$ (\ref{widgg}), which would be suppressed. If the particles carrying QCD color are heavier by factors of 2 or more, the hierarchy of decay rates  changes, as $\chi_3^2/\chi_\gamma^2\sim10^{-2}-10^{-4}$. In this case, decays to gluons no longer dominate the partial widths $\Gamma^{(8)}$ for the $C$-even states. Also, the regime for $m_0$ and $yM$ in which $\Gamma^{(8)}$ dominates over $\Gamma^{(6)}$ is somewhat reduced. Without any detailed computation,  we may infer from figures \ref{bounds1} and \ref{bounds2} that the decays of the $0^{++}$ v-glueball, with its large branching fraction to off-shell Higgs boson, will not be much affected. For $m_{0^-}<2m_W$, the dominant decay mode of the $0^{-+}$ v-glueball is now $\gamma\gamma$ with a branching ratio of $90\%$ with the decays into $gg$ and $Z\gamma$ accounting for the remaining $10\%$. In the high mass range $m_{0^-}>2m_W$, the $0^{-+}$ v-glueball decays dominantly into $WW$ with a branching ratio of $\sim 70\%$ followed by  the decays into  $ZZ$, $\gamma\gamma$ and $Z\gamma$ with branching ratios of $20\%$, $5\%$ and $5\%$, respectively. Also, the lifetime of the $0^{-+}$ v-glueball is longer. For the $2^{++}$ and $2^{-+}$ v-glueballs, the decays into photon pairs typically dominate for $m_{2^{\pm +}}-m_{0^{\pm +}}<m_H$, while the Higgs-radiative transitions $2^{\pm +}\to 0^{\pm +}h$ can dominate for $m_{2^{\pm +}}-m_{0^{\pm +}}>m_H$. The decays of the v-glueballs in the $C$-odd sector are largely unaffected. We will comment further  in the conclusions and in our LHC study~\cite{pg2}, but suffice it to say that the fact that the annihilation into photon pairs may dominate the lifetimes of some v-glueballs (the $0^{-+}$, $2^{++}$, $2^{-+}$) has important consequences for experimental searches.

\subsection{Uncertainties in the v-glueball matrix elements}
In another respect, we should emphasize that even though our  estimates are in general robust, they are subject to significant uncertainties, due especially to the many unknown v-glueball matrix elements. Although lattice computations are at present not available, it is of interest to see what one might expect once lattice computations are incorporated. Of course, the decays of the $0^{++}$ v-glueball will remain unchanged, since its branching ratios are independent of the decay constant ${\bf F_{0^{+}}^s}$. The same is true for the $1^{+-}$ and $0^{-+}$ v-glueballs. On the contrary, the decays of the $2^{++}$ and $1^{--}$ v-glueballs depend on ratios of the unknown  non-perturbative matrix elements,
\be
r_{2^+}=\frac{{\bf M_{2^{++}0^{++}}^S} m_{0}}{{\bf F_{2^{++}}^T}} \;\;\;\;\;\; r_{1^-}=\frac{{\bf M_{1^{--}1^{+-}}^S} m_{1^{+-}}}{{\bf M_{1^{--}0^{++}}^\Omega}}.
\ee
Since there is no reason for the form factors ${\bf M_{2^{++}0^{++}}^S}$ and ${\bf M_{1^{--}1^{+-}}^S}$  to be significantly suppressed or enhanced at $k^2=m_H^2$, the ratios $r_{2^+}$ and $r_{1^-}$  are probably of order ${\cal O}(1)$ (actually somewhat smaller\footnote{Using standard large-$n_v$ counting rules and taking into account explicit factors of $1/\sqrt{n_v}$ in the coupling constant, one obtains $r_{2^+}\sim{\cal O}(1/n_v)$ and $r_{1^-}\sim{\cal O}(1/\sqrt{n_v})$; see (\ref{approx-me}). }, especially at large $n_v$).
To illustrate the possible range of impact lattice computations may lead to, we can simply make the rescaling $yM\to yM \sqrt{r_{2^+}}$ and $yM\to yM \sqrt{r_{1^-}}$ in figure \ref{bounds1}. 
One can see that when $r_{2^+}$ is varied by a factor of $10$ in both directions around  $r_{2^+}\sim 1$ for $yM\sim 1\TeV$, the ratio $BR^{(6)}_{2^{+}}$ changes by around $10-20\%$, and the dominant mode of the $2^{++}$ v-glueball is still $gg$. A similar conclusion is obtained for the $2^{-+}$ v-glueball.
For the $1^{--}$ v-glueball, in contrast to the $2^{++}$ and $2^{-+}$, the branching ratios are more uncertain. The reason is that the contributions from $D=6$ and $D=8$ operators are comparable in strength, so a small enhancement in one of the matrix elements can lead to
a large effect on the branching ratios. For $r_{1^{-}}\sim 0.1$, the photon-radiative decay gets enhanced very greatly and could dominate the branching ratio in most of the mass range, while $r_{1^{-}}\sim 10$ tends to enhance the Higgs-radiative decay. Therefore, additional lattice computations would be needed to distinguish between these two modes.
Fortunately, it turns out that the impact of precise lattice computations on the phenomenology of the lightest v-glueballs $0^{++}$, $2^{++}$, $0^{-+}$ and $2^{-+}$, which, as  will be argued in our LHC study, are probably the most copiously produced, is in general very mild.

\section{Other extensions}
\setcounter{equation}{0}
In this section, we consider two simple extensions that may alter the phenomenology of v-glueballs in a number of ways, with a special emphasis on the $0^{-+}$ and $1^{+-}$ v-glueballs. Firstly, we analyze   the two Higgs doublet model (2HDM), which is the  simplest extension of the standard model Higgs sector, and secondly, we study the possibility  of $CP$ violation in the theory and its implications for v-glueball decays.
We will not attempt to systematically compute all the decay rates, but simply point out a few salient features of these extensions.

\subsection{2HDM}

As in the MSSM, we consider the SM with two Higgs doublets $H_u$ and $H_d$, where $H_u$ only couples to up-type quarks and neutrinos and $H_d$ only couples to down-type quarks and leptons. In this  model, there are five physical Higgs bosons: two charged scalars ($H^{\pm}$); two neutral scalars ($H$ and $h$); and a neutral $CP$-odd scalar ($A$). The presence of the $CP$-odd scalar $A$ is of particular interest.  Since the $A$  is $CP$-odd, a gauge invariant $CP$-conserving coupling to the v-gluons must be of the form  $\epsilon^{\mu\nu\alpha\beta} \tr {\cal F}_{\mu\nu}{\cal F}_{\alpha\beta} A\Phi$, with $\Phi=1,H,h$. 
Because of this coupling, new possible decays of the pseudoscalar $0^{-+}$ state emerge in the 2HDM. These are
\begin{eqnarray} \label{0p2hdm}
0^{-+}\;&\to&\; A^*\to\; ff,\\
 0^{-+}\;&\to& \;HA,\\
 0^{-+}\;&\to& \;hA\\
  0^{-+}\;&\to&\; 0^{++}A. \label{0p2hdm2}
\end{eqnarray}
For all other v-glueballs, there are new contributions to their decays in the 2HDM, but the decay pattern as described in sections 3 and 5 remains basically unchanged. For the $0^{++}$ v-glueball, the new channels include $0^{++}\to hh, Hh,AA$  as well as the off-shell Higgs decay $0^{++}\to h^*$. 
For all other states, new decay modes arise from the processes $\Theta_\kappa\to \Theta_{\kappa'}\phi$, where now $\phi = H,h,A$. Recall that in the single-Higgs model, the lowest-dimension operators that can induce decays of the $1^{+-}$ v-glueball arise at mass dimension eight. A similar conclusion applies to the $1^{+-}$ v-glueball in the context of the 2HDM. Therefore, while the $CP$-odd scalar $A$ makes a drastic change on the $0^{-+}$, it does not affect the $1^{+-}$ v-glueball.

To estimate the effect of the $CP$-odd scalar $A$ on the branching ratios, we will need to find the effective Lagrangian.
 Using the basic results of section 2, the effective Higgs couplings to  v-gluons, induced at one-loop by the $X$ particles, is given by
\begin{multline}\label{2hdm}
{\cal L}^{(6)}= \frac{\alpha_v }{ 3 \pi\, M^2} \,\left[y_u^2 \,H_u^\dagger H_u + y_d^2\, H_d^\dagger H_d\right] \,\tr {\cal F}_{\mu\nu} {\cal F}^{\mu\nu}  +\\+\frac{\alpha_v }{ 2 \pi\, M^2}\,\left[y_d^2\left(\frac{H_d-H_d^*}{2i}\right)^\dagger \left(\frac{H_d+H_d^*}{2}\right)+ y_u^2\left(\frac{H_u-H_u^*}{2i}\right)^\dagger \left(\frac{H_u+H_u^*}{2}\right)   \right] \,\tr { \cal F}_{\mu\nu} \tilde {\cal  F}^{\mu\nu}.
\end{multline}
 Here $y_u$ and $y_d$ represent the couplings of the $X$ particles to the Higgs doublets $H_u$ and $H_d$, respectively. 
The couplings of the physical Higgs bosons to v-gluon pairs are readily obtained by including the appropriate mixing angle factors
\begin{multline}\label{2hdm}
{\cal L}^{(6)}= \frac{\alpha_v }{ 6 \pi\, M^2} \,\left( \xi_{HH} H^2 +\xi_{hh} h^2+ \xi_{H h} H h + \xi_{AA} A^2 \right) \,\tr {\cal F}_{\mu\nu} {\cal F}^{\mu\nu}  + \frac{\alpha_v \, v_H}{ 2 \pi\, M^2} \,\xi_A \, A\,\tr { \cal F}_{\mu\nu} \tilde { \cal F}^{\mu\nu} \\+\frac{\alpha_v \,v_H }{ 3 \pi\, M^2} \,\left( \xi_H H +\xi_h h \right) \,\tr {\cal F}_{\mu\nu} {\cal F}^{\mu\nu} +\frac{\alpha_v }{ 4 \pi\, M^2} \,\left( \xi_{AH} A H +\xi_{Ah}A h \right) \,\tr { \cal F}_{\mu\nu} \tilde {\cal  F}^{\mu\nu}
\end{multline} 
where the coeffients $\xi_{ij}$ and $\xi_i$ are given by
\be\nonumber \label{2hdmcouplings}
 \xi_{HH}= y_u^2 s_\alpha^2+y_d^2 c_\alpha^2 \;\;\;\; \xi_{hh}= y_u^2 c_\alpha^2+y_d^2 s_\alpha^2  \;\;\;\; \xi_{Hh}= 2 s_\alpha c_\alpha (y_u^2-y_d^2)\ee
 \be \nonumber
  \xi_{AH}=y_u^2 s_\alpha c_\beta-y_d^2 c_\alpha s_\beta \;\;\;\;  \xi_{Ah} = y_u^2 c_\alpha c_\beta +y_d^2 s_\alpha s_\beta \;\;\;\; \xi_{AA}= y_u^2 c_\beta^2 -y_d^2 s_\beta^2 \ee
 \be
  \xi_H= y_u^2 s_\alpha s_\beta + y_d^2 c_\alpha c_\beta \;\;\;\; \xi_h= y_u^2 c_\alpha s_\beta - y_d^2 s_\alpha c_\beta \;\;\;\;\xi_A = s_\beta c_\beta (y_u^2-y_d^2)
 \ee
 where  $\alpha$  the mixing angle  and the ratio of the vacuum expectation values $\tan \beta=v_u/v_d$. The combination $v_d^2+v_u^2$ is fixed by the electroweak scale $v_H=\sqrt{v_d^2+v_u^2}= 246\GeV$. 

 The decay rates  can be readily extracted from our general results in section 3 by replacing $y^2$ with the appropriate couplings  in (\ref{2hdmcouplings}).
 For the scalar $0^{++}$ state, the  decay rates for the off-shell Higgs decay $0^{++}\to H^*/h^*\to \zeta\zeta$  can be read from (\ref{wid0}) through the substitution
\be
\frac{y^2}{m_0^2-m_H^2}\to \frac{\xi_H}{m_0^2-m_H^2}+\frac{\xi_h}{m_0^2-m_h^2},
\ee
while  the  rate for the two-body decays $0^{++}\to HH, \,hh,\, Hh,\, AA$  is obtained from (\ref{wid0hh}) by replacing
 \be
 y^2\left(1+\frac{3 m_Z^2}{2(m_0^2-m_H^2)} \right) \to \xi_{ij} +\frac{\xi_H g_{Hij}}{m_0^2-m_H^2}+\frac{\xi_h g_{hij}}{m_0^2-m_h^2},
 \ee
 where $i,j= HH,Hh,hh,AA$ and $g_{ijk}$ are the  cubic self-couplings in the 2HDM. Likewise, the width of the decays $\Theta_\kappa\to \Theta_\kappa' \Phi_i$, $\Phi_i=H,h,A$, can be obtained from (\ref{jjh}) and (\ref{3body}) by substituting $y^2 \to \xi_i$. Finally, we summarize
 the rates for the new decay modes of the $0^{-+}$ v-glueball (\ref{0p2hdm}-\ref{0p2hdm2}),  
\be \label{wid0pp}
\Gamma_{0^{-+}\rightarrow f \bar f} = \left(\frac{\xi_A \,\tan\beta\, \,v_H  \,F_{0^+}^P}{2 \pi M^2 (m_A^2-m_0^2)}   \right)^2 \Gamma^{SM}_{h \rightarrow f\bar f}(m_{0^-}^2)
\ee
\be
\Gamma_{0^{-+}\rightarrow 0^{++}A} =\frac{1}{16\pi  m_{0^-}} \left(\frac{{\bf M_{0^-0^+}^P} \xi_{A} }{2\pi M^2 }\right)^2  \left[g(m_H^2,m_A^2;m_{0^-}^2) \right]^{1/2}
\ee
\be
\Gamma_{0^{-+}\rightarrow AH} =\frac{1}{16\pi  m_{0^-}} \left(\frac{ F_{0^-}^P}{2\pi M^2 }\right)^2 \left( \xi_{AH} +\frac{3 \xi_A} {2(m_{0^-}^2-m_A^2)}\right)^2 \left[g(m_H^2,m_A^2;m_{0^-}^2) \right]^{1/2}
\ee
\be
\Gamma_{0^{-+}\rightarrow Ah} =\frac{1}{16 \pi  m_{0^-}} \left(\frac{ F_{0^-}^P}{2\pi M^2 }\right)^2 \left(\xi_{Ah}+\frac{3 m_Z^2 \xi_A}{2(m_{0^-}^2-m_A^2)}\right)^2 \left[g(m_h^2,m_A^2;m_{0^-}^2) \right]^{1/2}.
\ee

The expressions above are clearly very model dependent. First, notice that for  $y_u\simeq y_d$, $\xi_{A}$ is very small, and the decay rate of the $0^{-+}$ v-glueball is suppressed. On the other hand, for $y_d\simeq 0$, a quick check shows that this rate, whose   ratio to the $0^{++}$ width is (for small masses $m_0 < m_h$)
\be
\frac{\Gamma_{0^{-+}}}{\Gamma_{0^{++}}}\simeq 0.5 \left( s_\alpha \frac{m_A^2}{m_H^2}+ c_\alpha \frac{m_A^2}{m_h^2}\right)^{-2},
\ee
is not negligible. This ratio may be anywhere from $\sim 0.1$ up to about $4$ depending on the mixing angle and the masses of the scalars. Unless the $0^{-+}$ decay rate is unduly suppressed, the lifetimes of the  $0^{++}$ and  $0^{-+}$ v-glueballs are in general  within one order of magnitude from each other.

\subsection{Models of explicit $CP$ violation}

It is of interest to see what one might expect once  $CP$-violation effects are incorporated in the theory, in the context of a single-Higgs model. For example, parity-violating interaction terms   can be induced in the effective Lagrangian if we assume that the  couplings of the heavy particles $X_r$  to the Higgs boson are complex. In this case, a $P$-odd $C$-even interaction is generated of the form $ H^\dagger H \,\tr {\cal F}_{\mu\nu} \tilde {\cal F}^{\mu\nu}$. As in the 2HDM, 
 this interaction  drastically changes  $0^{-+}$ decays, since without it, the $0^{-+}$ v-glueball  would decay via dimension-eight operators, suffering an extra suppression.
 The $0^{-+}$ v-glueball can then decay via $0^{-+}\to h^*$, producing any of the kinematically-allowed final states of the Higgs boson. 

Similarly, one can contemplate interactions  that explicitly break C-invariance. Interestingly enough,  the lowest-dimension operators of this kind (e.g. $H^\dagger H d_{abe }f_{cde } {\cal F}^a_{\mu\nu} {\cal F}^b_{\mu\nu}  {\cal F}^c_{\alpha\beta} {\cal F}^d_{\alpha\beta}$) arise at dimension ten and, therefore, their effects are extremely suppressed by extra powers of $1/M$. The dominant decay modes of the $1^{+-}$ v-glueball are then induced by dimension-eight operators, even if we allow for $CP$-violation.

In the rest of this section, then, we will focus on $P$-violating but $C$-conserving interactions.
To estimate the decay widths, let us consider the following Lagrangian, in  two-component notation,
 \be
\lag_{mass}=  y_d  X_q H^\dagger X_{d}^c +  y_d  X_q^c H X_{d} +h.c. \label{yuk-complex}
\ee
where $y_d$ is a complex Yukawa coupling. 
By integrating out the $X_q$ and $X_d$ particles, we obtain the following  dimension-six operators\footnote{The effective Lagrangian (\ref{efflag2}) can be evaluated either by following the procedure displayed in appendix A  or, equivalently,  in the limit of vanishing Higgs momentum by taking derivatives of the v-gluon self-energy and the axial anomaly:\be\nonumber{\cal L}^{(6)}= H^\dagger H/v_H^2 \, (y_d^2 \partial^2/\partial y_d^2 + {y_d^*}^2\partial^2/\partial {y_d^*}^2) \,{\cal L},\ee where \be\nonumber {\cal L} = (\alpha_v/6\pi) \ln (\Lambda_{UV}^2/\mathrm{det} M^\dagger M) \,\tr {\cal F}_{\mu\nu} {\cal F}^{\mu\nu} + (\alpha_v/2\pi i) \ln (\mathrm{det} M/\mathrm{det} M^\dagger ) \,\tr {\cal F}_{\mu\nu} \tilde {\cal F}^{\mu\nu}\ee with $M=\left(\begin{array}{cc} M& y_d v_H\\ y_d v_H& M \end{array}\right)$.}:

\be\label{efflag2}
{\cal L}^{(6)}= \frac{\alpha_v \,(y^2-\tilde y^2)}{ 3 \pi\, M^2} \, H^\dagger H \,\tr {\cal F}_{\mu\nu} {\cal F}^{\mu\nu} + \frac{\alpha_v \,(2 \,y\,\tilde y)}{  \pi\, M^2} \, H^\dagger H \,\tr {\cal F}_{\mu\nu} \tilde {\cal F}^{\mu\nu}.
\ee
where $y= Re\, y_d$ and $\tilde y =Im \,y_d$.
Here $\tilde {\cal F}_{\mu\nu} =(1/2)\epsilon_{\mu\nu\lambda\sigma}{\cal F}^{\mu\nu}$ denotes the dual of the field strength tensor. The first term on the right-hand side of  (\ref{efflag2}) is the same operator that we had already found in (\ref{efflag}). The  second term of (\ref{efflag2}) is new and violates $P$. As mentioned above, the $P$-odd interaction allows the pseudoscalar $0^{-+}$ state to decay into SM particles
 via $s$-channel Higgs-boson exchange $0^{-+}\rightarrow h^{*}\to \zeta \zeta$, where 
$\zeta$  denotes  a standard model particle. 
The width of the decay is given by
\be \label{wid0p}
\Gamma_{0^{-+}\rightarrow \zeta \zeta} = \left(\frac{2 y\tilde y \,v_H \, \alpha_v \,F_{0^-}^P}{\pi M^2 (m_H^2-m_{0^-}^2)}   \right)^2 \Gamma^{SM}_{h \rightarrow \zeta \zeta}(m_{0^-}^2)
\ee
where $F_{0^-}^P \equiv \alpha_v \,\langle
0|\tr {\cal F}_{\mu\nu} \tilde {\cal F}^{\mu\nu} |0^{++}\rangle$ is the $0^{-+}$ decay constant. 
The same operator also induces the decay $0^{-+}\to 0^{++} h$, with partial width
\be
\Gamma_{0^{-+}\rightarrow 0^{++}h} =\frac{1}{16\pi  m_{0^-}} \left(\frac{2y\tilde y\,v_H\,\alpha_v  \, {\bf M_{0^-0^+}^P}  }{\pi \,M^2 }\right)^2  \left[g(m_H^2,m_{0^+}^2;m_{0^-}^2) \right]^{1/2}
\ee
where now $ {\bf M_{0^-0^+}^P}  $ is the transition matrix. This decay is  phase-space suppressed  for $m_{0^{-+}}- m_{0^{++}}<m_h$, but may be quite significant  for $m_{0^{-+}}- m_{0^{++}}>m_h$ when the radiated Higgs boson is onshell. For example, for $m_h\sim 100\GeV$ we obtain $BR(0^{-+}\to 0^{++} h)\sim 0.3$.

In the SM, the most stringent limits on the size of the $CP$ violation effects come from experimental limits on the electric dipole moment of the neutron. The same experimental constraints also allow us to place limits on the $CP$-violating operator in (\ref{efflag2}). To see this,  let us recall that  the $X$ particles can also carry QCD  color.  Then, the following  dimension-six operators are induced:
\be\label{efflag3}
{\cal L}^{(6)}= \frac{\alpha_s \,(y^2-\tilde y^2)}{ 3 \pi\, M^2} \, H^\dagger H \,\tr {G}_{\mu\nu} {G}^{\mu\nu} + \frac{\alpha_s \,(2 \,y\,\tilde y)}{  \pi\, M^2} \, H^\dagger H \,\tr { G}_{\mu\nu} \tilde { G}^{\mu\nu}.
\ee
where now $G_{\mu\nu}$ denotes the SM gluon field strength tensor and $\tilde G_{\mu\nu}$ its dual. The $CP$ violating term in (\ref{efflag3}) with the Higgs field replaced by its vacuum expectation value contributes to the $\theta$ angle. We assume that this contribution to the $\theta$ angle is removed by whatever  mechanism that solves the strong $CP$ problem in QCD. The next operators in the expansion around the Higgs VeV contain one or two powers of the Higgs field. Their contribution to the electric dipole moment of the neutron can be estimated using Naive Dimensional Analysis~\cite{Georgi1,Georgi2}. Following the analysis of ~\cite{Manohar:2006gz,Weinberg:1989dx}, we estimate the neutron electric dipole moment  as
\be
d_n = \frac{e}{2\pi f_\pi} \frac{\alpha_s \,(2 \,y\,\tilde y)}{ 4 \pi\, M^2}\, \frac{(2\pi f_\pi)^2}{(4\pi)^2}  \sim 10^{-25}{\rm e-cm}\,\times\, \frac{(2 y \tilde y) \,\alpha_s}{8 \pi^2} \,\times\, (1\TeV/M^2).
\ee
where $2\pi f_\pi \simeq 1190\MeV$ is the chiral-symmetry-breaking scale.	
Combining this result with the current experimental   bound on $d_n$, we derive a limit on $ 2 y\tilde y$
\be
|2 y\tilde y| < 10 \times (M/1\TeV)^2.
\ee
Consequently, values for $|y|$, $|\tilde y|$  in the range $0.01-1$ are consistent with the existing experimental limits.

\section{Conclusions}

In this work, we have investigated a particularly challenging hidden valley scenario with a broader class of couplings to SM particles than those considered in ~\cite{Juknevich:2009ji}. In particular, we have focused on the effect of dimension-six operators  by which the  hidden sector interacts with the standard model  through the Higgs sector. 

The resulting   v-glueball phenomenology  is fairly complex in this scenario, but there are  some simple features.
In particular, we find the  following interesting signatures:
\begin{itemize}
\item Decays of the $0^{++}$ v-glueball through $h^*$, producing any of the kinematically-allowed Higgs final states such as $b\bar b$, $\tau^{+}\tau^-$, $WW$, etc. 
 \item  Multiple Higgs boson emission, from cascade decays $\Theta_\kappa\to \Theta_{\kappa'} h$, or annihilations $0^{++}\to hh$.
\item Due to the diversity of v-glueball states, and the presence of operators of different mass dimension, the lifetimes can vary at least over 5 or 6 orders of magnitude, for any given choice of parameters (see figure \ref{lifetimes}). This has two immediate consequences: (i) Displaced vertices are quite common, which could potentially serve as a discovery channel, and (ii) v-glueballs can be as light as a few GeV's and still be visible.
\item For sufficiently small v-glueball masses, a different opportunity arises in the form of non-standard Higgs decays such as $h\to \Theta_\kappa \Theta_{\kappa'}$, with branching ratio to v-glueballs of order $10^{-3}-10^{-2}$.
\end{itemize}

In addition, other possible final states include $gg$, $\gamma\gamma$, and radiative transitions with emission of a photon, all of which are induced by dimension-eight operators~\cite{Juknevich:2009ji}. The  rare diphoton channel is particularly interesting because, thanks to its moderate QCD background, it may serve as the discovery channel for the v-glueballs at the LHC. 

We should nevertheless emphasize that some of our results are subject to significant theoretical and numerical uncertainties. In this respect,  it would be interesting to know the spectrum of pure-Yang-Mills theory for gauge groups other than $SU(3)$, as well as the various v-glueball matrix elements that arise. With enough motivation, such as a hint for discovery, these could in principle be determined by additional lattice computations. If the v-sector gauge group is not  $SU(n_v)$, some of the v-glueballs may not be present; for instance, for $SO(n_v)$ or $Sp(n_v)$ gauge groups the $C$-odd sector is absent or heavy. However, as explained in \cite{Juknevich:2009ji},  the lightest $C$-even v-glueballs are expected to be present in any pure-gauge theory, with similar production and decay channels, so at least for them, the basic features of a pure-gauge hidden valley are expected to be retained. We also  have not considered higher-order corrections to the decay rates, and they should be  taken into account when precise predictions are required.

We have seen that the operators considered in this paper are not heavily constrained by current experimental searches or precision electroweak data, thus
leaving ample parameter space to be explored by the next generation of hadron collider experiments.  Application of our results for phenomenological studies, particularly as relevant for the LHC, will be carried out in a companion paper~\cite{pg2}. We expect that detection should be feasible, if the mass $M$ of the $X$ particles is small enough to give a reasonable large cross-section, and $\Lambda_v$ is large enough to ensure the v-glueballs decay inside the detectors.
Decays to bottom quarks and  gluons often dominate, but they suffer from a huge multijet background. 
At the Tevatron and at the LHC, the v-glueballs are most likely to be found in searches for diphoton  resonances in events with 2 photons plus jets or with 3 or more photons and possibly additional jets, or in searches for displaced decays to jet pairs, $W/Z$ pairs, or photon pairs.  A potentially novel signature, which could be currently searched for at the Tevatron, arises in the form of two b-tagged jets plus diphoton events.  Contrary to the models of \cite{H01,H02,H03}, the branching ratio to  $b\bar b\gamma\gamma$ in our model can be ${\cal O}(1)$, since the $b\bar b$ and $\gamma\gamma$ final states originate in different resonances, leading to a potentially discoverable signal despite the background of $2j+2\gamma$. Altogether, these signatures provide an interesting opportunity to search for new physics at the Tevatron and at the LHC, which motivates further theoretical and experimental study.

\bigskip
{ \Large \bf Acknowledgments}

\smallskip \smallskip

I would like to thank Matt Strassler for suggesting this problem and for many helpful and enlightening discussions.
I would  also like to thank Adam Falkowski, Yuri Gershtein, Jessie Shelton, Scott Thomas and Gonzalo Torroba for helpful conversations, and to Stephen Martin for pointing out an error in a previous version of equation (\ref{decay_constants}). This work was supported by the Department of Energy under grant DE-FG02-96ER40949.


\appendix
\renewcommand{\theequation}{A-\arabic{equation}}
  \setcounter{equation}{0}  
  \section{Computation of ${\cal L}^{(6)}_{eff}$}  

\subsubsection*{Definitions}
In the following the momenta of the initial gluons ($p_1,p_2$) and final Higgs bosons ($k_1,k_2$) are all chosen to be incoming. We make use of the following invariants  
\be
s = (p_1+p_2)^2 \;\;\;\; t=(p_1+k_4)^2 \;\;\;\; u=(p_1+k_3)^2
\ee
 obeying the condition
\be
s+t+u= k_3^2+k_4^2
\ee
where we have assumed that the gluons are massless, $p_1^2=p_2^2=0$.

\subsubsection*{Pasarino-Veltman functions $C_0$ and $D_0$}

The computation of the Feynman  graphs is performed using the Passarino-Veltman decomposition of the one-loop tensor integrals~\cite{passarino}. For our results below we will make use of the 3 and 4 point scalar functions $C_0$ and $D_0$. The explicit expressions are rather lengthy and have to be handled numerically. However, one can derive an integral representation
\begin{multline}
C_0(r_1,r_2,s,M_1,M_2,M_3) = \int_{0}^1 \int_{0}^x \left[-r_2 x^2 - 
   r_1 y^2 + (-s + r_1 + r_2) x y+ \right.\\ \left.+ (M_3 - M_2 + r_2) x + (M_2 - M_1 + s - 
      r_2) y - M_3 \right]^{-1} dy dx
\end{multline}

\begin{multline}
D_0(r_1,r_2,r_3,r_4,s,t,M_1,M_2,M_3,M_4)=\int_{0}^1 \int_{0}^x \int_{0}^y
\left[-r_3 x^2 - r_2 y^2  - 
   r_1 z^2 +\right. \\ \left. +(-t + r_1 + r_4) x y + (s + t - r_2 - r_4) x z + (-s + r_1 + 
      r_2) y z + (M_4 - M_3 + r_3) x + \right.\\ +\left. (M_3 - M_2 + t - r_3) y + (M_2 - M_1 + 
      r_4 - t) z - M_4\right]^{-2} \,dz \, dy \,dx
\end{multline}

\subsubsection*{Fermion loop}
\begin{figure}[ht]
\begin{center}

\epsfxsize=16cm \epsffile{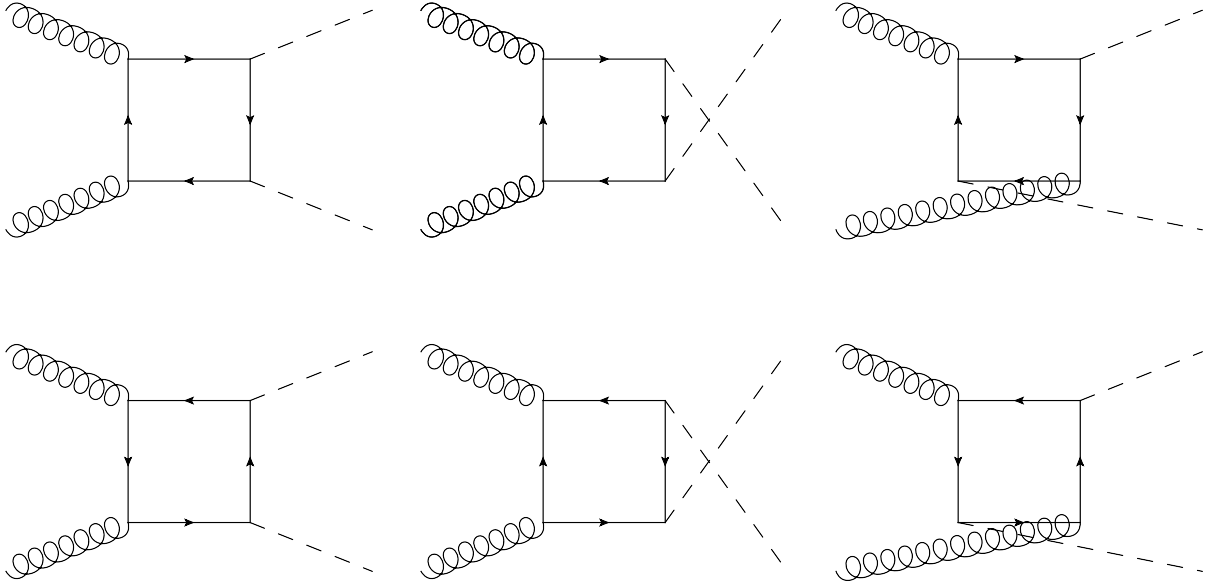}

\end{center}

\caption{Graphs contributing to $gg \to HH$}
\label{brhiggs1}
\end{figure}

The amplitude for the fusion of two massless v-gluons into a pair of Higgs boson fields arises at lowest order in perturbation theory from a loop of heavy particles. The relevant graphs are depicted in figure \ref{brhiggs1}. We assume that there is a pair of vector-like fermion fields, a doublet and a singlet of $SU(2)_L$, transforming under the fundamental representation of $SU(n_v)$. The doublet and singlet have masses $M$ and $m$, respectively. Gauge invariance constrains the amplitude to be
\be
{\cal M}_\square = \frac{y^2 g_v^2}{8 \pi^2}\left( p_1^\nu p_2^\mu -p_1\cdot p_2 g^{\mu\nu}\right) F_\square \; \epsilon_\mu^1  \epsilon_\nu^2 \delta_{ab} + \cdots
\ee
where $\epsilon^1_\mu$, $\epsilon_\mu^2$ are the polarization vectors of the incoming v-gluons.
The form factor $F_\square$  is given by:
\be
F_\square = f(s,t,u,k_3^2,k_4^2,m,M) +f(s,t,u,k_4^2,k_3^2,m,M)+ (M\leftrightarrow m)
\ee
where
\begin{multline}
f(s,t,u,k_3^2,k_4^2,m,M)=\\
-\frac{1}{s^2} (k_3^2+k_4^2-2(M+m)^2) \left[(k_4^2-u)C_0(0,k_4^2,u,M^2,M^2,m^2)+(k_4^2-t)C_0(0,k_4^2,t,m^2,m^2,M^2)\right.\\\left.+(k_3^2-u)C_0(0,u,k_3^2,m^2,m^2,M^2)+(k_3^2-t)C_0(0,t,k_3^2,u,M^2,M^2,m^2) \right]\\
+ 8 \frac{m^2}{s} C_0(0,0,s,m^2,m^2,m^2)\\
-2 \frac{m^2 }{s}(-2 (m+M)^2  + k_3^2 + k_4^2 + \frac{M}{m} s) \left[D_0(0,0,k_3^2,k_4^2,s,t,m^2,m^2,m^2,M^2)\right. \\ \left.+D_0(0,0,k_4^2,k_3^2,s,u,m^2,m^2,m^2,M^2)\right]+\\
\frac{1}{s^2} \left( 2 m^4 s+4 m^3 M s+m^2 (s (4 M^2-k_4^2)+2 k_3^3 k_4^2-k_3^2 (2 k_4^2+s))+2 m M (2 M^2 s-2 k_3^2 (k_4^2-t)\right.\\ \left.+2 k_4^2 t-s^2-2 s t-2 t^2)+(2 M^2-k_3^2-k_4^2) (s M^2+k_3^3 k_4^2-k_3^2 k_4^2) \right.\\ \left.+
s (2 m^2 +2 M^2 -k_3^2 -k_4^2 )p_t^2 \right) D_0(0,k_4^2,0,k_3^2,t,u,m^2,m^2,M^2,M^2) +4
\end{multline}
with $p_t = 2 (p_1 k_3)(p_2 k_3)/(p_1 p_2)-k_3^2$.

\subsubsection*{Large fermion mass limit and effective Lagrangian}
The form factor can be evaluated by taking the limit $M,m \gg s,t,u,k_3^2,k_4^2$. At leading order in the mass splitting of the heavy fermions $M\approx m$,  the form factor reads
\be
F_\square = \frac{8}{3 M m} +{\cal O} (s/M^2) .
\ee
The matrix element for $g_ag_b\to H H$ can be obtained in perturbation theory from the following non-renormalizable interaction
\be \label{eff}
 \lag_{eff}^{(6)}=\frac{y^2\alpha_v}{3 \pi M m}  \, H^\dagger H \tr{{\cal F}_{\mu\nu}{\cal F}^{\mu\nu}}.
\ee

\section{Classification of v-sector operators}

The classification of the operators ${\cal O}^{(d)}_v$ in terms of irreducible representations of the Lorentz group was carried out in \cite{Jaffe}. (See also ~\cite{Juknevich:2009ji} for a review.) In this appendix we summarize the operators which are relevant for our work.
Here $S$, $P$, $T$, $L$ and $\Omega^{(1,2)}$ label the decomposition of each operator ${\cal O}_v^{(d)}$ into irreducible representations of the Lorentz group. Their expressions are shown in tables~\ref{dim4} and~\ref{dim6} along with the  $J^{PC}$ states that each operator can create acting on the vacuum.
\begin{table}[h]
\begin{center}
\begin{tabular}[c]{|cc|}\hline \Trule\Brule
\ Operator ${\cal O}^{\xi}_v$&$J^{PC}$
\\ \hline \Trule
 $S= \tr \Fc_{\mu\nu}\Fc^{\mu\nu}$ & $0^{++}$
\\
 $P=\tr \Fc_{\mu\nu}\tilde \Fc^{\mu\nu}$ & $0^{-+}$
\\
 $T_{\alpha\beta}=\tr  \Fc_{\alpha\lambda}\Fc^{~\lambda}_{\beta} -\frac14\, g_{\alpha\beta} S$ & $2^{++}$, $1^{-+}$, $0^{++}$
\\
  $ L_{\mu\nu\alpha\beta}=\tr \Fc_{\mu\nu}\Fc_{\alpha\beta} -\frac12 \, (g_{\mu\alpha}T_{\nu\beta}+g_{\nu\beta}T_{\mu\alpha}-g_{\mu\beta}T_{\nu\alpha}-g_{\nu\alpha}T_{\mu\beta})
$ & $2^{++}$, $2^{-+}$ \\ \Brule
 $-\frac{1}{12}\,(g_{\mu\alpha}g_{\nu\beta}-g_{\mu\beta}g_{\nu\alpha})S +\frac{1}{12}\,\epsilon_{\mu\nu\alpha\beta}P$   &
\\ \hline
\end{tabular}
\end{center}
\caption{\small The dimension $d=4$ operators, and the states that can
be created by these operators \cite{Jaffe}. We denote $\tilde
\Fc_{\mu\nu}=\frac12\, \epsilon_{\mu\nu\alpha\beta}\Fc^{\alpha\beta}$.}
\label{dim4}
\end{table} From now on, we denote the
operators ${\cal O}_v^\xi$, where $\xi$ runs over different
irreducible operators $\xi= S, P, T, L, \cdots$.
\begin{table}[ht]
\begin{center}
\begin{tabular}[c]{|cc|}\hline\Trule\Brule
\ Operator ${\cal O}^{\xi}_v$ &$J^{PC}$
\\ \hline \Trule
 $\Omega^{(1)}_{\mu\nu}= \tr \Fc_{\mu\nu}\Fc_{\alpha\beta}\Fc^{\alpha\beta}$ & $1^{--}$, $1^{+-}$ \\
\Brule  $\Omega^{(2)}_{\mu\nu}= \tr \Fc_\mu^{\alpha} \Fc^{\beta}_{\alpha}\Fc_{\beta\nu}$& $1^{--}$, $1^{+-}$
\\ \hline
\end{tabular}
\end{center}
\caption{\small The important $d=6$ operators. 
The states that can be created by these operators are shown \cite{Jaffe}.}
\label{dim6}
\end{table}

\section{The form factors ${\cal M}^{(i)}_{JJ'} $}
 In addition to the momenta $p$ and $q$, we introduce  a pair of tensors $\epsilon_{\mu_1,\cdots,\mu_J}(p)$ and   $\tilde \epsilon_{\mu_1,\cdots,\mu_{J'}}(q)$ to represent the polarizations of the spin-J and spin-J' states respectively. It will be convenient to define "reduced" polarization tensors by contracting some of the indices with the vectors $p$ and $q$,
\be
\epsilon_{\mu_1,\cdots,\mu_{J-K}}(p) =\epsilon_{\mu_1,\cdots,\mu_J}(p) q^{\mu_{J-K+1}} \cdots q^{\mu_J} \;\;\;\;\; \tilde \epsilon_{\mu_1,\cdots,\mu_{J'-K'}}(q) =\tilde \epsilon_{\mu_1,\cdots,\mu_J}(q) p^{\mu_{J'-K'+1}}\cdots p^{\mu_{J'}}.
\ee
The invariant form factors are constructed from the reduced polarization tensors and the vectors $p$ and $q$. For the simplest $J\to 0$ case, the only  form factor is obtained when all the indices of the polarization tensor are fully contracted with $q$,
\be
\epsilon(p) = \epsilon_{\mu_1,\cdots,\mu_J}(p) \,q^{\mu_{0}} \cdots q^{\mu_J}.
\ee
This gives the matrix element
\be
\ME{0,q}{S}{J,p}\, = \,  {\bf M^S_{0 J}} \, \epsilon_{\mu_1,\cdots,\mu_J}(p) \,q^{\mu_{0}} \cdots q^{\mu_J}
\ee
where now ${\bf M^S_{0 J}}  $ is the transition matrix which depends on the   transferred momentum.
For the more complex $J\to J'=1$ case, there are  two form factors parametrizing the matrix element,
\be
 \epsilon_{\mu,\mu_2\cdots,\mu_J}(p) \,q^{\mu_{2}} \cdots q^{\mu_J} \,\tilde\epsilon^{\mu}(q),\;\;\;\; \;\;\epsilon_{\mu_1,\mu_2\cdots,\mu_J}(p) \,q^{\mu_{1}} \cdots q^{\mu_J} \,\tilde\epsilon_{\nu}(q) \,p^{\nu}.
\ee
In the reduced notation, these are simply $\epsilon_\mu \,\tilde\epsilon^\mu=$ and $\epsilon\, \tilde \epsilon$. In addition, we can contract the polarization tensors with a Levi-Civita tensor as follows,
\be
\epsilon^{\mu\nu\rho\sigma} \epsilon_{\mu}(p)   \, p_\rho \,q_\sigma \tilde \epsilon_\nu(q)=\epsilon^{\mu\nu\rho\sigma} \epsilon_{\mu,\mu_2\cdots,\mu_J}(p)  \,q^{\mu_{2}} \cdots q^{\mu_J} \, p_\rho \,q_\sigma \tilde \epsilon_\nu(q)
\ee
Collecting all terms, we obtain the following matrix element
\begin{multline}
\ME{1,q}{S}{J,p} =  {\bf M^S_{1 J}}\epsilon_{\mu,\mu_2\cdots,\mu_J}(p) \,q^{\mu_{2}} \cdots q^{\mu_J} \,\tilde\epsilon^{\mu}(q)+
{\bf M^{S'}_{1 J}} \epsilon_{\mu_1,\mu_2\cdots,\mu_J}(p) \,q^{\mu_{1}} \cdots q^{\mu_J} \,\tilde\epsilon_{\nu}(q) \,p^{\nu}+\\
{\bf M^{S''}_{1 J}} \epsilon^{\mu\nu\rho\sigma} \epsilon_{\mu,\mu_2\cdots,\mu_J}(p)  \,q^{\mu_{2}} \cdots q^{\mu_J} \, p_\rho \,q_\sigma \tilde \epsilon_\nu(q)
\end{multline}

We next turn to the more complicated matrix element for the $J\to J'=2$ transition. Firstly, we define the auxiliary tensors
\be
\epsilon_{\mu\nu}(p)= \epsilon_{\mu,\nu,\mu_3\cdots,\mu_J}(p) \,q^{\mu_{3}} \cdots q^{\mu_J}, \;\;\;\;\epsilon_{\mu}(p)= \epsilon_{\mu,\mu_2\cdots,\mu_J}(p) \,q^{\mu_{2}} \cdots q^{\mu_J}
\ee
and
\be
\tilde \epsilon_{\mu}(q)= \tilde \epsilon_{\mu\nu}(q) \,p^{\nu}.
\ee
From $\epsilon_{\mu\nu}(p)$, $\epsilon_{\mu}(p)$, $\tilde \epsilon_{\mu\nu}(q)$  and $\tilde \epsilon_{\mu}(q)$ we can form the following three invariants
\be
\epsilon_{\mu\nu}(p)\,\tilde \epsilon_{\mu\nu}(q),\;\;\;\; \epsilon_{\mu}(p)\,\tilde \epsilon_{\mu}(q),\;\;\;\;(\epsilon_{\mu}(p) \,q^\mu)\,(\tilde \epsilon_{\nu}(q)\,p^\nu)
\ee
In addition, contracting with the Levi-Civita tensor gives two more invariants,
\be
\epsilon^{\mu\nu\rho\sigma} \epsilon_{\mu}(p)   \, p_\rho \,q_\sigma \tilde \epsilon_\nu(q),\;\;\;\;\epsilon^{\mu\nu\rho\sigma} \epsilon_{\mu\lambda}(p)   \, p_\rho \,q_\sigma \tilde \epsilon_\nu^{\,\,\,\lambda}(q).
\ee
Therefore, there are five form factors in the $J\to J'=2$ case. The general rule to find all the form factors in the case $J\to J'$ case is now clear. Without loss of generality we can assume $J>J'$. First we construct the reduced spin-J  tensor with $J'$ indices,
\be
\epsilon_{\mu_1,\cdots,\mu_{J'}}(p) =\epsilon_{\mu_1,\cdots,\mu_J}(p) q^{\mu_{J-J'}} \cdots q^{\mu_J} 
\ee
so that we have two tensors with $J'$ indices. Then we form  the invariants
\be
\epsilon_{\mu_1,\cdots,\mu_{J'}}(p) \tilde \epsilon^{\mu_1, \cdots \mu_{J'}}(q),\;\cdots,\; \epsilon_\mu(p) \,\tilde \epsilon^{\mu}(q), \; \epsilon(p)  \,\tilde \epsilon(q),
\ee
for a total of $J'+1$ invariant form factors. Secondly, there are contractions with the Levi-Civita tensor as follows
\be
\epsilon_{\mu\nu\rho\sigma} p^\rho q^\sigma \epsilon_{\mu,\mu_2,\cdots,\mu_{J'}}(p) \tilde \epsilon^{\nu\mu_2, \cdots \mu_{J'}}(q),\;\cdots,\; \epsilon_{\mu\nu\rho\sigma} p^\rho q^\sigma \epsilon_\mu(p) \,\tilde \epsilon^{\nu}(q)
\ee
giving $J'$ additional invariants. Hence, the total number of Lorentz invariant form factors is $2 J'+1$ for the transitions $J\to J'$, $J'<J$.   
To see that this contemplate all the possibilities, we can apply angular momentum counting rules. By composing the spin $J$ and spin $J'$ states we obtain total spin $|J-J'|,\cdots, J+J'$. For a scalar operator, these have to be combined with total angular momentum $L= |J-J'|,\cdots, J+J'$ in order to give total spin 0. So for $J'<J$ we get $2 J'+1$ form factors associated with the different values of $L$.
\begin{table}[ht]
\begin{center}
\begin{tabular}[c]{|c|c|c|c|c|c|}\hline \Trule\Brule 
$i$&1 &2&3&4&5 
\\ \hline\Trule\Brule
${\cal M}^{(2,i)}_{\alpha\beta\rho\sigma}$ &$g_{\alpha\rho}\,g_{\beta\sigma}$ &$g_{\alpha\rho}\, \hat q_\beta\, \hat p_\sigma$&$\hat q_\alpha \hat p_\rho\, \hat q_\beta\, \hat p_\sigma$&$g_{\beta\rho}\,\epsilon_{\alpha\sigma z w}\,\hat p^z \,\hat q^w$& $\hat q_\beta \,\hat p_\rho\,\epsilon_{\alpha\sigma zw}\,\hat p^z \,\hat q^w$\\ \hline\Trule\Brule
${\cal M}^{(1,i)}_{\alpha\rho}$ &$g_{\alpha\rho}$&$q_\alpha \,\hat p_\rho$&$\epsilon_{\alpha\rho z w}\,\hat p^{\,z} \hat q^{\,w}$& -&-\\\hline
\end{tabular}
\end{center}
\caption{\small The auxiliary tensors ${\cal M}^{(J,i)}$. Here $\hat q_\alpha = q_\alpha/\sqrt{q^2}$ and $\hat p_\alpha = p_\alpha/\sqrt{p^2}$. }
\label{auxtensors}
\end{table}

To summarize, we show the matrix elements up to spin 3 that are needed to compute the decays of the v-glueballs in figure \ref{spectrum}. In terms of the auxiliary tensors shown in table \ref{auxtensors}, the matrix elements read,
\be\label{m32}
{\cal M}^{(i)}_{32} = \,\hat q_\gamma\,\epsilon^{\alpha\beta\gamma}(p) \,\tilde \epsilon^{\,\rho\sigma}(q)\,{\cal M}^{(2,i)}_{\alpha\beta\rho\sigma}\,
\ee
\be
{\cal M}^{(i)}_{31} = \,\hat q_\gamma\,\hat q_\beta\,\epsilon^{\alpha\beta\gamma}(p) \,\tilde \epsilon^{\,\rho}(q)\,{\cal M}^{(1,i)}_{\alpha\rho}
\ee
\be
{\cal M}_{30} = \hat q_\alpha\,\hat q_\beta\,\,\hat q_\gamma\, \epsilon^{\alpha\beta\gamma}(p)\, 
\ee

\be
{\cal M}^{(i)}_{22} = \,\epsilon^{\alpha\beta}(p) \,\tilde \epsilon^{\,\rho\sigma}(q)\,{\cal M}^{(2,i)}_{\alpha\beta\rho\sigma}
\ee
\be
{\cal M}^{(i)}_{21} = \,\hat q_\beta\,\epsilon^{\alpha\beta}(p) \,\tilde \epsilon^{\,\rho}(q)\,{\cal M}^{(1,i)}_{\alpha\rho}
\ee
\be
{\cal M}_{20} = \hat q_\alpha\,\hat q_\beta\,\, \epsilon^{\alpha\beta}(p) 
\ee

\be
{\cal M}^{(i)}_{11} = \,\epsilon^{\alpha}(p) \,\tilde \epsilon^{\,\rho}(q)\,{\cal M}^{(1,i)}_{\alpha\rho}
\ee
\be\label{m10}
{\cal M}_{10} = \hat q_\alpha\,\, \epsilon^{\alpha}(p) 
\ee
where we denote $\hat q_\alpha = q_\alpha /\sqrt{q^2}$. For the sake of brevity, we have not shown the seven matrix elements that correspond to the $3 \to 3$ transition. These can be obtained along the same line that led to (\ref{m32})-(\ref{m10}).



\begin{thebibliography}{99}
\bibitem{SZ} M.~J.~Strassler and K.~M.~Zurek,
  Phys.\ Lett.\  B {\bf 651} (2007) 374
  [arXiv:hep-ph/0604261].

\bibitem{HV2}
  M.~J.~Strassler and K.~M.~Zurek,
  Phys.\ Lett.\  B {\bf 661}, 263 (2008)
  [arXiv:hep-ph/0605193].

\bibitem{HV3}
  M.~J.~Strassler,
  arXiv:hep-ph/0607160.

\bibitem{HVun}
  M.~J.~Strassler,
  arXiv:0801.0629 [hep-ph].


\bibitem{HVWis}
  T.~Han, Z.~Si, K.~M.~Zurek and M.~J.~Strassler,
  JHEP {\bf 0807}, 008 (2008)
  [arXiv:0712.2041 [hep-ph]].


\bibitem{hvstudy1}
  M.~J.~Strassler,
  arXiv:0806.2385 [hep-ph].


\bibitem{TwinHiggs}
  Z.~Chacko, H.~S.~Goh and R.~Harnik,
  Phys.\ Rev.\ Lett.\  {\bf 96}, 231802 (2006)
  [arXiv:hep-ph/0506256].

\bibitem{FoldedSUSY}
  G.~Burdman, Z.~Chacko, H.~S.~Goh and R.~Harnik,
  JHEP {\bf 0702}, 009 (2007)
  [arXiv:hep-ph/0609152].
  
\bibitem{hvdarkmatter}K.~M.~Zurek,
  arXiv:0811.4429 [hep-ph];
 J.~March-Russell, S.~M.~West, D.~Cumberbatch and D.~Hooper,
  JHEP {\bf 0807} (2008) 058
  [arXiv:0801.3440 [hep-ph]];
 N.~Arkani-Hamed and N.~Weiner,
  JHEP {\bf 0812}, 104 (2008)
  [arXiv:0810.0714 [hep-ph]];
  A.~E.~Nelson and C.~Spitzer,
  arXiv:0810.5167 [hep-ph].

\bibitem{string hv} 
  R.~Blumenhagen, M.~Cvetic, P.~Langacker and G.~Shiu,
  Ann.\ Rev.\ Nucl.\ Part.\ Sci.\  {\bf 55}, 71 (2005)
  [arXiv:hep-th/0502005].

\bibitem{Juknevich:2009ji}
  J.~E.~Juknevich, D.~Melnikov and M.~J.~Strassler,
  JHEP {\bf 0907}, 055 (2009)
  [arXiv:0903.0883 [hep-ph]].




\bibitem{Morningstar}
 C.~J.~Morningstar and M.~J.~Peardon,
 Phys.\ Rev.\  D {\bf 60} (1999) 034509
 [arXiv:hep-lat/9901004].

\bibitem{Morningstar2} Y.~Chen {\it et al.},
 Phys.\ Rev.\  D {\bf 73} (2006) 014516
 [arXiv:hep-lat/0510074].
\bibitem{Shifman scalar}
 V.~A.~Novikov, M.~A.~Shifman, A.~I.~Vainshtein and V.~I.~Zakharov,
 Nucl.\ Phys.\  B {\bf 165}, 67 (1980).


\bibitem{Shifman pseudoscalar}
 V.~A.~Novikov, M.~A.~Shifman, A.~I.~Vainshtein and V.~I.~Zakharov,
 Nucl.\ Phys.\  B {\bf 165}, 55 (1980);
V.~A.~Novikov, M.~A.~Shifman,
A.~I.~Vainshtein and V.~I.~Zakharov,
 Phys.\ Lett.\  B {\bf 86}, 347 (1979)
 [JETP Lett.\  {\bf 29}, 594.1979\ ZFPRA,29,649 (1979\
ZFPRA,29,649-652.1979)].

\bibitem{MITbagmodel}
  J.~F.~Donoghue, K.~Johnson and B.~A.~Li,
  Phys.\ Lett.\  B {\bf 99} (1981) 416.

\bibitem{glueballs} J.~M.~Cornwall and A.~Soni,
  Phys.\ Lett.\  B {\bf 120} (1983) 431.

\bibitem{bagmodel} J.~Kuti,
  Nucl.\ Phys.\ Proc.\ Suppl.\  {\bf 73} (1999) 72
  [arXiv:hep-lat/9811021]

\bibitem{Loan}
  M.~Loan and Y.~Ying,
  Prog.\ Theor.\ Phys.\  {\bf 116}, 169 (2006)
  [arXiv:hep-lat/0603030].


\bibitem{glueballreview}
  V.~Mathieu, N.~Kochelev and V.~Vento,
  Int.\ J.\ Mod.\ Phys.\  E {\bf 18}, 1 (2009)
  [arXiv:0810.4453 [hep-ph]].

 
\bibitem{Faraggi:2000pv}
  A.~E.~Faraggi and M.~Pospelov,
  Astropart.\ Phys.\  {\bf 16}, 451 (2002)
  [arXiv:hep-ph/0008223].


\bibitem{Falkowski:2009yz}
  A.~Falkowski, J.~Juknevich and J.~Shelton,
  arXiv:0908.1790 [hep-ph].

 
\bibitem{Kribs:2009fy}
  G.~D.~Kribs, T.~S.~Roy, J.~Terning and K.~M.~Zurek,
  arXiv:0909.2034 [hep-ph].

\bibitem{Shifman charmonium}
 V.~A.~Novikov, L.~B.~Okun, M.~A.~Shifman, A.~I.~Vainshtein, M.~B.~Voloshin and V.~I.~Zakharov,
  Phys.\ Rept.\  {\bf 41}, 1 (1978).



\bibitem{Strassler:1992nc}
 M.~J.~Strassler,
 SLAC-PUB-5978


\bibitem{Groote}
 S.~Groote and A.~A.~Pivovarov,
 Eur.\ Phys.\ J.\  C {\bf 21}, 133 (2001)
 [arXiv:hep-ph/0103313].


 
\bibitem{hunter}
For a review on Higgs Physics, see: J. F. Gunion, H. E. Haber, G. Kane, S. Dawson, {\it The Higgs Hunter's guide} (Addison-Wesley, Reading Mass., 1990)
 
\bibitem{Jaffe}
 R.~L.~Jaffe, K.~Johnson and Z.~Ryzak,
 Annals Phys.\  {\bf 168} (1986) 344.



\bibitem{Okun} L.~B.~Okun,
  JETP Lett.\  {\bf 31} (1980) 144
  [Pisma Zh.\ Eksp.\ Teor.\ Fiz.\  {\bf 31} (1979) 156],
  Nucl.\ Phys.\  B {\bf 173} (1980) 1.

\bibitem{quinn}
S.~Gupta and H.~R.~Quinn,
  Phys.\ Rev.\  D {\bf 25} (1982) 838.




\bibitem{KLN}
  J.~Kang, M.~A.~Luty and S.~Nasri,
  JHEP {\bf 0809}, 086 (2008)
  [arXiv:hep-ph/0611322].

  J.~Kang and M.~A.~Luty,
  JHEP {\bf 0911}, 065 (2009)
  [arXiv:0805.4642 [hep-ph]].





\bibitem{cdf1}
  A.~Abulencia {\it et al.}  [CDF Collaboration],
  Phys.\ Rev.\  D {\bf 75}, 112001 (2007)
  [arXiv:hep-ex/0702029].

\bibitem{cdf2}
  T.~Aaltonen {\it et al.},
  arXiv:0910.5170 [hep-ex].

R.~Culbertson {\it et al.},[CDF Collaboration], Public note

\bibitem{Aaltonen:2008dm}
  T.~Aaltonen {\it et al.}  [CDF Collaboration],
  Phys.\ Rev.\  D {\bf 78}, 032015 (2008)
  [arXiv:0804.1043 [hep-ex]].


\bibitem{Kribs:2007nz}
  G.~D.~Kribs, T.~Plehn, M.~Spannowsky and T.~M.~P.~Tait,
  Phys.\ Rev.\  D {\bf 76}, 075016 (2007)
  [arXiv:0706.3718 [hep-ph]].

\bibitem{Abazov:2009ik}
  V.~M.~Abazov {\it et al.}  [D0 Collaboration],
  Phys.\ Rev.\ Lett.\  {\bf 103}, 071801 (2009)
  [arXiv:0906.1787 [hep-ex]].
\bibitem{Abazov:2008zm}
  V.~M.~Abazov {\it et al.}  [D0 Collaboration],
  Phys.\ Rev.\ Lett.\  {\bf 101}, 111802 (2008)
  [arXiv:0806.2223 [hep-ex]].


\bibitem{Lynn:1985sq}
  B.~W.~Lynn, G.~Penso and C.~Verzegnassi,
  Phys.\ Rev.\  D {\bf 35}, 42 (1987).



\bibitem{Peskin:1991sw}
  M.~E.~Peskin and T.~Takeuchi,
  Phys.\ Rev.\  D {\bf 46}, 381 (1992).


\bibitem{pg2}
J.~E.~Juknevich, D.~Melnikov and M.J.~Strassler, in preparation.


\bibitem{Georgi1}
  H.~Georgi, A.~Manohar and G.~W.~Moore,
  Phys.\ Lett.\  B {\bf 149}, 234 (1984).
\bibitem{Georgi2}
  H.~Georgi and L.~Randall,
  Nucl.\ Phys.\  B {\bf 276}, 241 (1986).


\bibitem{Manohar:2006gz}
  A.~V.~Manohar and M.~B.~Wise,
  Phys.\ Lett.\  B {\bf 636}, 107 (2006)
  [arXiv:hep-ph/0601212].

\bibitem{Weinberg:1989dx}
  S.~Weinberg,
  Phys.\ Rev.\ Lett.\  {\bf 63}, 2333 (1989).

\bibitem{H01}
  S.~Chang, P.~J.~Fox and N.~Weiner,
  Phys.\ Rev.\ Lett.\  {\bf 98}, 111802 (2007)
  [arXiv:hep-ph/0608310].

\bibitem{H02}
  R.~Dermisek and J.~F.~Gunion,
  Phys.\ Rev.\  D {\bf 77}, 015013 (2008)
  [arXiv:0709.2269 [hep-ph]].
\bibitem{H03}
  B.~A.~Dobrescu, G.~L.~Landsberg and K.~T.~Matchev,
  Phys.\ Rev.\  D {\bf 63}, 075003 (2001)
  [arXiv:hep-ph/0005308].
\bibitem{passarino}
  G.~'t Hooft and M.~J.~G.~Veltman,
  Nucl.\ Phys.\  B {\bf 153}, 365 (1979).

\end{thebibliography}
\end{document}